\documentclass[twocolumn]{aastex62}

\newcommand{\rev}[1]{{\color{purple}#1}}

\newcommand{\todo}{\ifmmode \text{\color{purple}\Huge{\(\bullet\)}} \else {\color{purple}{\Huge$\bullet$}}\fi}
\newcommand{\REFS}{\todo \rev{REFS}}

\newcommand{\mstar}{M_{\star}}
\newcommand{\msun}{M_{\odot}}
\newcommand{\mbh}{M_{\rm BH}}
\newcommand{\nh}{N_{\rm H}}

\newcommand{\akari}{\textit{AKARI}}
\newcommand{\wise}{\textit{WISE}}

\newcommand{\iras}{\textit{IRAS}}
\newcommand{\galex}{\textit{GALEX}}
\newcommand{\spitzer}{\textit{Spitzer}}
\newcommand{\iso}{\textit{ISO}}
\newcommand{\hst}{\textit{HST}}
\newcommand{\um}{\mathrm{\mu m}}

\newcommand{\mdotbh}{\dot{M}_\mathrm{BH}}
\newcommand{\lbat}{L_\mathrm{14-150 keV}}
\newcommand{\lsun}{L_\odot}

\newcommand{\lbol}{L_\mathrm{AGN,bol}}
\newcommand{\ledd}{L_{\rm Edd}}
\newcommand{\eddington}{\lambda_\mathrm{Edd}}

\newcommand{\mdotcool}{\dot{M}_{\mathrm{cool}}}

\newcommand{\lambdaedd}{\lambda_\mathrm{Edd}}

\usepackage{natbib}
\usepackage{amsmath}
\usepackage{multirow} 
\usepackage{textgreek}
\usepackage{xcolor}
\usepackage{booktabs} 
\usepackage{ifthen} 
\usepackage{comment}
\received{April 3, 2022}
\accepted{September 16, 2022}
\submitjournal{ApJ}

\shorttitle{BH growth in Galaxy cluster}
\shortauthors{Fukuchi et al.}

\DeclareUnicodeCharacter{2212}{-} 
\begin{document}

\title{H1821+643: The most X-ray and infrared luminous AGN in the \textit{Swift}/BAT survey in the process of rapid stellar and supermassive black hole mass assembly}

\correspondingauthor{Hikaru Fukuchi, Kohei Ichikawa}
\email{hikaru.fukuchi@astr.tohoku.ac.jp, k.ichikawa@astr.tohoku.ac.jp}

\author[0000-0001-7557-6854]{Hikaru Fukuchi}

\affil{Astronomical Institute, Tohoku University, Aramaki, Aoba-ku, Sendai, Miyagi 980-8578, Japan}

\author[0000-0002-4377-903X]{Kohei Ichikawa}
\affil{Astronomical Institute, Tohoku University, Aramaki, Aoba-ku, Sendai, Miyagi 980-8578, Japan}
\affil{Frontier Research Institute for Interdisciplinary Sciences, Tohoku University, Sendai 980-8578, Japan}

\author[0000-0002-2651-1701]{Masayuki Akiyama}
\affil{Astronomical Institute, Tohoku University, Aramaki, Aoba-ku, Sendai, Miyagi 980-8578, Japan}

\author{Claudio Ricci}
\affil{Núcleo de Astronomía de la Facultad de Ingeniería, Universidad Diego Portales, Av. Ejército Libertador 441, Santiago 22, Chile}
\affil{Kavli Institute for Astronomy and Astrophysics, Peking University, Beijing 100871, People's Republic of China}

\author{Sunmyon Chon}
\affil{Astronomical Institute, Tohoku University, Aramaki, Aoba-ku, Sendai, Miyagi 980-8578, Japan}

\author{Mitsuru Kokubo}
\affil{Astronomical Institute, Tohoku University, Aramaki, Aoba-ku, Sendai, Miyagi 980-8578, Japan}
\affil{
Department of Astrophysical Sciences, Princeton University, Princeton, New Jersey 08544, USA
}
\author{Ang Liu}
\affil{
Max Planck Institute for Extraterrestrial Physics,
Giessenbachstrasse 1, 85748 Garching, Germany
}

\author{Takuya Hashimoto}
\affil{
Tomonaga Center for the History of the Universe (TCHoU), Faculty of Pure and Applied Sciences, University of Tsukuba, Tsukuba, Ibaraki 305-8571, Japan
}
\author{Takuma Izumi}
\affil{
National Astronomical Observatory of Japan, 2-21-1 Osawa, Mitaka, Tokyo 181-8588, Japan
}
\affil{Department of Astronomical Science, The Graduate University for Advanced Studies, SOKENDAI, 2-21-1 Osawa, Mitaka, Tokyo 181-8588, Japan}



\begin{abstract}

H1821+643 is the most X-ray luminous non-beamed AGN of $\lbat = 5.2\times 10^{45}$~erg~s$^{-1}$ in the \textit{Swift}/BAT ultra-hard X-ray survey and it is also a hyper-luminous infrared (IR) galaxy $L_\mathrm{IR} = 10^{13.2}
L_\odot$ residing in the center of a massive galaxy cluster, which is a unique environment achieving the rapid mass assembly of black holes (BH) and host galaxies in the local universe.
We decompose the X-ray to IR spectral energy distribution (SED) into the AGN and starburst component using the SED fitting tool \verb|CIGALE|-2022.0 and show that H1821+643 consumes a large amount of cold gas ($\dot{M}_\mathrm{con}$) with star-formation rate of $\log ( \mathrm{SFR}/\msun~\mathrm{yr}^{-1}) = 3.01 \pm 0.04$ and BH accretion rate of $\log (\mdotbh/\msun~\mathrm{yr}^{-1}) = 1.20 \pm 0.05$. 
This high $\dot{M}_\mathrm{con}$ is larger than the cooling rate ($\dot{M}_\mathrm{cool}$) of the intra-cluster medium (ICM), $\dot{M}_\mathrm{con}/\dot{M}_\mathrm{cool} \gtrsim 1$, which is one to two order magnitude higher than the typical value of other systems, indicating that H1821 provides the unique and extreme environment of rapid gas consumption.
We also show that H1821+643 has an efficient cooling path achieving from $10^7$~K to $10^2$~K thanks to [O~{\sc{i}}] 63 $\um$, which is a main coolant in low temperature range ($10^4$~K to $10^2$~K) with a cooling rate of $\mdotcool =3.2\times 10^5\ \mathrm{\msun ~yr^{-1}}$, and the star-forming region extends over 40 kpc scale.

\end{abstract}

\keywords{galaxies: SMBH growth: active --- galaxies: nuclei ---
quasars: general}

\section{Introduction}\label{sec:intro}

\begin{figure*}
\begin{center}
\includegraphics[width=0.48\textwidth]{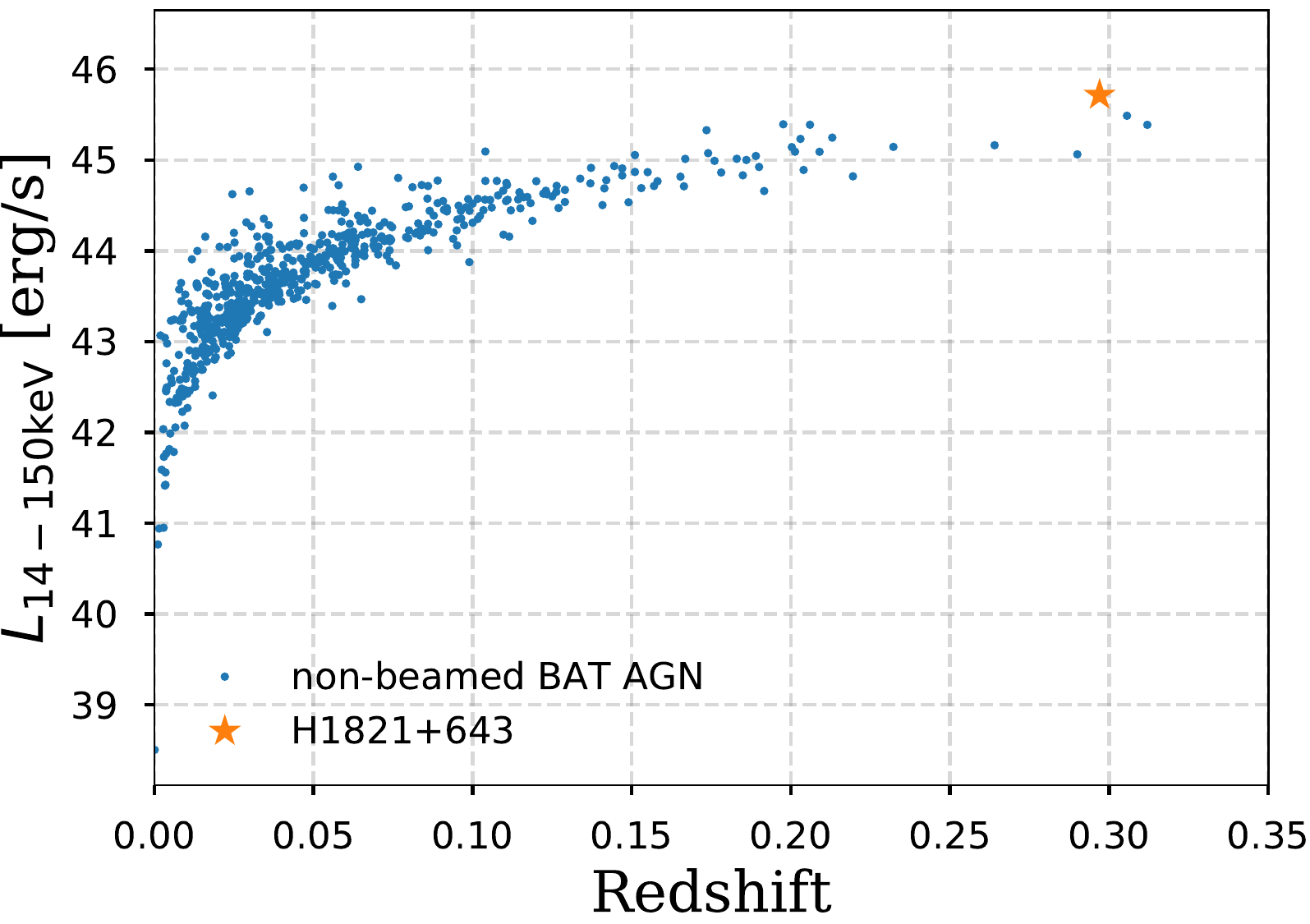}~
\includegraphics[width=0.48\textwidth]{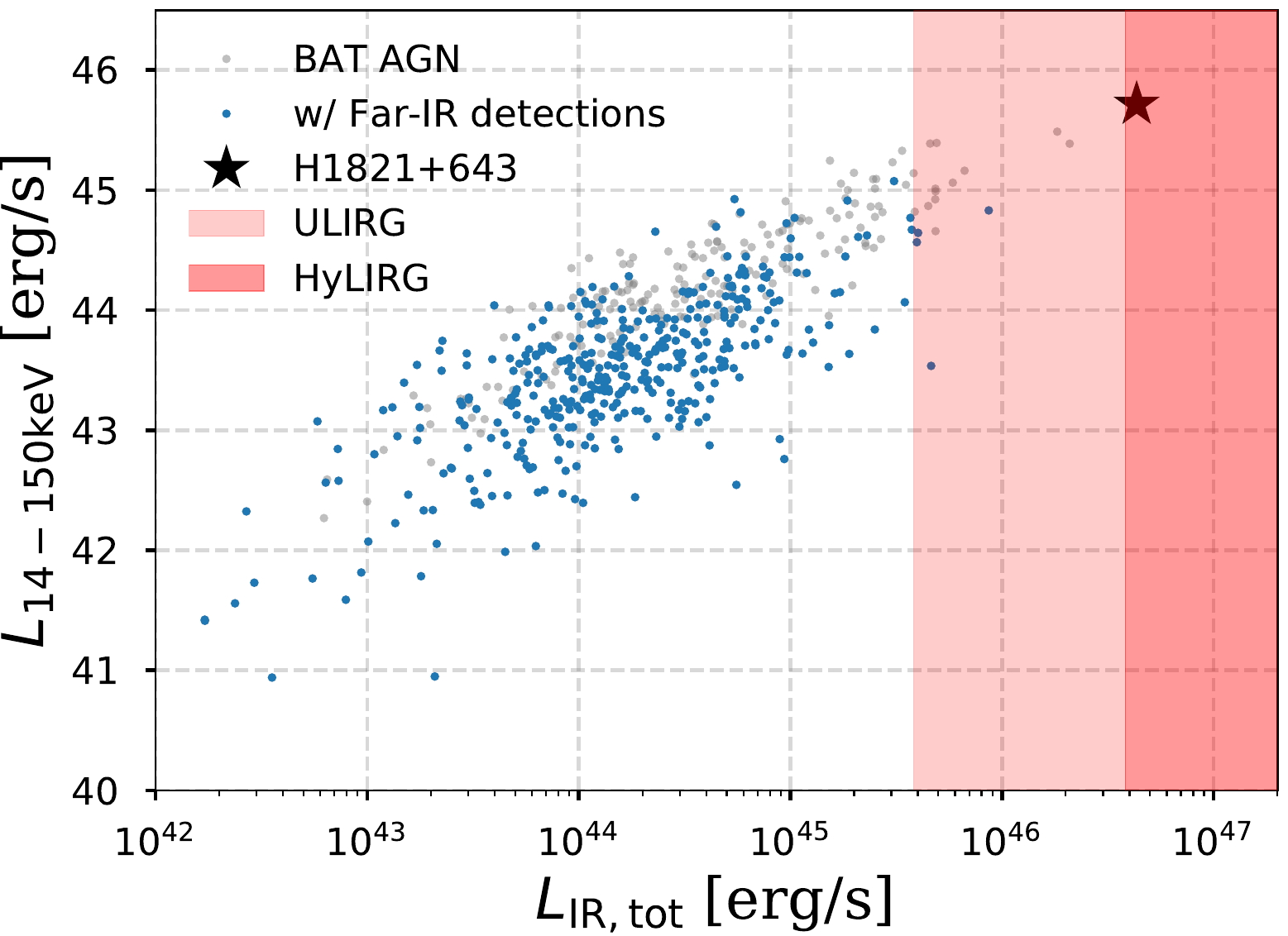}~
\caption{
(Left) Luminosity of the 70-month \textit{Swift}/BAT AGN sample as a function of redshift at $z<0.4$. H1821+643 is shown with the star (orange). The 14--150~keV luminosity is taken from \cite{ric17}. 
(Right) 14--150~keV luminosity and the total infrared luminosities (1--1000~$\mu$m) taken from \cite{ich19}.
}\label{fig:L_vs_z}
\end{center}
\end{figure*}

How supermassive black holes (SMBHs) in the local universe acquired their mass is one of the fundamental open questions in modern astronomy.
Active galactic nuclei (AGN), especially the high luminosity ones, are the best laboratories to understand the mass growth of SMBHs, 
since they are in a rapidly mass accreting phase $\mdotbh = L_\mathrm{AGN}/\eta c^2$,
releasing large amounts of gravitational energy
as radiation \citep[e.g.,][]{sol82,yu02}.
The accretion through this AGN phase finally leads to the achievable maximum mass limit of $\mbh \sim 10^{10.5} \ \msun$ \citep{net03,mcc11,kor13,tra14,jun15,ina16,ich17b}.


In the local universe, 
ultra-hard ($E>10$~keV) X-ray observations
are one of the most reliable methods for identifying
AGN in a complete way without worrying on the surrounding gas thanks to the strong penetration power against absorption \citep[e.g.,][]{ric15,ric17} and its very low contamination from other objects such as stellar-emission \citep[e.g.,][]{min12}.
Among the currently available surveys, 
\textit{Swift}/BAT survey provides the
most sensitive X-ray all-sky survey in the 14--195~keV band, reaching a 5$\sigma$ flux limit of (1.0--1.3)$\times 10^{-11}$~erg~s$^{-1}$~cm$^{-2}$ 
after the 70-month time integration \citep{bau13}
and a deeper 5$\sigma$ flux limit of (7.2--8.4)$\times 10^{-12}$~erg~s$^{-1}$~cm$^{-2}$ in the 105-month catalog \citep{oh18}.
Thanks to the intensive follow-up observations
by BAT AGN Spectroscopic Survey
\citep[BASS;][]{kos17,ric17,lam17,pow18,ich19},
it enables us to access the multi-wavelength properties
including the spectral energy distributions (SEDs), and several key physical quantities such as BH masses, X-ray luminosities, all of which are crucial to understand the BH growth during the AGN phase.

Among the AGN detected by \textit{Swift}/BAT survey, 
H1821+643 ($z=0.297$; hereinafter called ``H1821'', Swift ID; 967), 
whose optical coordinates are RA$=$18:21:57.312 and Dec$=$+64:20:36.24, 
is the most luminous AGN in the 14--150~keV band
except the beamed sources\footnote{Here, we removed the beamed sources (blazars) since their obtained 14--150~keV luminosities are
boosted and therefore their intrinsic values are difficult to estimate.} at $z<0.4$ as shown in the left panel of Figure~\ref{fig:L_vs_z}. 
The luminosity reaches $\lbat = 5.2\times 10^{45}$~erg~s$^{-1}$,
and the detailed X-ray spectral fitting shows an unobscured spectra with a very low column density of $\nh \le 10^{20}$~cm$^{-2}$ \citep{ric17}. The absorption corrected 2-10~keV luminosity is $L_\mathrm{2-10keV} = 5.8\times 10^{45}$~erg~s$^{-1}$
\citep{ric17}. 
This value is equivalent to the bolometric luminosity of $\lbol = 2.9 \times 10^{47}$~erg~s$^{-1}$ assuming the  luminosity dependent bolometric correction of 50 
for the AGN with $L_\mathrm{2-10keV} \sim 6\times 10^{45}$~erg~s$^{-1}$ \citep{vas07}.
The estimated BH mass of $\mbh$ = $3.9 \times 10^9 \ \msun$ \citep{kos17} is obtained from the optical spectra utilizing the H$\beta$ emission lines \citep[e.g.,][]{tra12}. The Eddington ratio of
$\eddington=0.59$ ($\lambdaedd \equiv \lbol / \ledd$, where $\ledd$ is Eddington luminosity $\ledd \simeq 1.26 \times 10^{38} (\mbh/\msun)$~erg~s$^{-1}$) is almost reaching to the Eddington limit.

Interestingly, H1821 was also detected in the multiple far-IR (FIR) bands by \textit{AKARI}/FIS and even by a shallower sensitivity IR survey of \textit{IRAS}/PSC \citep{ich19}.
The right panel of Figure~\ref{fig:L_vs_z} shows that the integrated total IR luminosity
at 1--1000~$\mu$m, compiled by the BASS survey \citep{ich19}, also reaches
$L_\mathrm{IR,tot} = 4.6 \times 10^{46}$~erg~s$^{-1}$ $=1.1\times 10^{13} \lsun$. 
Therefore, H1821 is categorized as 
a hyper-luminous IR galaxy \citep[HyLIRG; ][]{row91,san96,far02}, and is also the most IR luminous AGN in the \textit{Swift}/BAT AGN catalog.
These two properties of bright X-ray and FIR emissions suggest that H1821 hosts a rapidly growing SMBH in the center, 
and the host galaxy is also plausibly actively star-forming.
This is indirect evidence for the availability 
of cold gas supply fueling both star-formation, and the central SMBH.  
The large cold gas supply in H1821 can be achieved thanks to its special environment.
H1821 is actually the brightest cluster/central galaxy (BCG) of a galaxy cluster \citep[][see Appendix~\ref{app:h1821}]{sch92,hal97,sax97,fan02,rus10, wal14,rey14}, which is seemingly in contradiction with the picture that most BCGs in galaxy clusters are already quenched in the local universe \citep{pet03}. \cite{rus10} conducted a deep Chandra X-ray observation of H1821,
and found that H1821 has a strong cool core cluster \citep[e.g.,][]{hud10} whose central cooling time is $\sim$~1~Gyr~\footnote{Here, it is usually called a cool-core cluster, if a cluster has a dense core with a radiative cooling time shorter than 7.7~Gyr~\citep[e.g.,][]{hud10}, which corresponds to the look-back time of $z=1$}.
Thus, this extreme gas feeding might be supplied by the gas cooling in the core of galaxy cluster with inward flow of material \citep[cooling flow: see][for a review]{fab94}, as seen in the Phoenix cluster at $z=0.6$, with the association of a massive molecular gas \citep{rus17}. 
Considering its properties mentioned above, H1821 is an ideal local laboratory to explore how galaxy clusters fed their BCGs and SMBHs in the high-$z$ Universe at $z \sim 1$.

In this paper, we revisit the multi-wavelength
properties of H1821 and estimate the gas consumption rate in the system, by decomposing the UV, optical and IR SED. The main goal of this work is to 
quantitatively assess the consumption rate
by the SMBH (accretion rate to BH; $\mdotbh$) and host galaxy (star-formation rate; SFR) simultaneously (Section~\ref{sec:fitting} and \ref{sec:result}) by decomposing the AGN and host galaxy component through the SED fitting. We also discuss the cooling path from hot cluster gas to the cold gas feeding the mass assembly of the central galaxy H1821 (Section~\ref{sec:gas} and \ref{sec:dis}). 

Throughout the paper, we adopt standard cosmological parameters ($H_0 = 70.0$~km~s$^{-1}$~Mpc$^{-1}$, $\Omega_\mathrm{M}=0.3,$ and $\Omega_\Lambda = 0.7$).

\section{Multi-wavelength SED of H1821+643}\label{sec:analysis}

We summarize here the multi-wavelength information
from the IR, optical, UV and to X-ray data to apply the SED fitting and to obtain the AGN and host galaxy properties of H1821.
The whole photometric bands and flux values of each survey are summarized in Tables \ref{tab:flux1}--\ref{tab:flux_xray}. 
The multi-wavelength observational properties of H1821 are summarized in the Appendix~\ref{app:h1821}.

\begin{deluxetable}{cccccccc}
\tabletypesize{\footnotesize}
\tablecolumns{6}
\tablewidth{0pt}
\tablecaption{Summary of H1821+643 UV-IR fluxes \label{tab:flux1}}
  \tablehead{
      \colhead{Catalog, Band} & \colhead{$\lambda_{\mathrm{center}}$} & \colhead{Band Width} &\colhead{Flux} & \colhead{PSF} & \colhead{Ref.} 
      \\
       \colhead{} & \colhead{($\mathrm{\mu m}$)} & \colhead{($\mathrm{\mu m}$)} & \colhead{(mJy)}
       & \colhead{(arcsec)} & \colhead{}
     }
\startdata
    $\galex$/FUV &        0.15 &        0.02 &        6.63 $\pm$ 0.01 &          4.5"   &       M \\
    $\galex$/NUV &        0.23 &        0.07 &        7.32 $\pm$ 0.01 &         6.0"    &        M \\
    Johnson $V$ &        0.55 &        0.06 &        9.11 
    &       --   &         R, F \\
    Johnson $R$ &        0.64 &        0.19 &        12.2 $\pm$ 0.4 &        --    &        R, O \\
    Johnson $I$ &        0.80 &        0.21 &        13.7 $\pm$ 0.9 &       --   &        R, O \\
     2MASS $J$&          1.25 &        0.21 &    11.90 $\pm$ 0.20 &              0.23" &         a \\
      2MASS $H$&          1.65 &        0.25 &    14.73 $\pm$ 0.29 &              0.33" &         a \\
      2MASS $Ks$ &          2.16 &        0.27 &    23.46 $\pm$ 0.37 &              0.43" &         a \\
      $\wise$ W1 &          3.4 &        0.8 &      52.0 $\pm$ 1.1 &               6.1" &         b \\
      $\wise$ W2 &          4.6 &        1.1 &      79.1 $\pm$ 1.5 &               6.4" &         b \\
      $\wise$ W3 &         12 &        9 &     160.1 $\pm$ 2.1 &               6.5" &         b \\
      $\wise$ W4 &         22 &        4 &     447.8 $\pm$ 7.9 &                12" &         b \\
     $\iras$/PSC &         25 &       11 &    395.2 $\pm$ 59.3 &               4.7' &         c \\
     $\iras$/PSC &         60 &       30 &  1128 $\pm$ 113 &               4.5' &         c \\
     $\iras$/PSC &        100 &       30 &  2159 $\pm$ 259 &               5.3' &         c \\
    $\akari$/IRC &          9.0 &        4.2 &     130.8 $\pm$ 3.7 &               5.5" &         d \\
    $\akari$/IRC &         18.0 &       11.1&     325.9 $\pm$ 9.6 &               5.7" &         d \\
    $\akari$/FIS &         90 &       24 &    634.8 $\pm$ 34.1 &                39" &         e \\
     $\spitzer$/IRS &         14.0 &     11--18 &     228.7 $\pm$ 3.0 &  --  &     f \\
 $\spitzer$/IRS &         21.4 &     18--26 &     385.3 $\pm$ 5.5 &  --     &  f \\
 $\spitzer$/IRS &         30.0 &     26--35 &    492.3 $\pm$ 7.0 &     --   & f \\
         $\iso$ &         92 &    80--100 &   1860 $\pm$ 322 & -- &       g \\
         $\iso$ &        108 &   100--120 &   2068 $\pm$ 325 & --  &      g \\
         $\iso$ &        126 &   120--140 &   1779 $\pm$ 343 & -- &       g \\
         $\iso$ &        151 &   140--160 &   1489 $\pm$ 251 & --  &      g \\
\enddata
\tablecomments{We show UV to FIR information of H1821+643 used in SED fitting. IR fluxes are obtained from NASA/IRSA: \url{https://irsa.ipac.caltech.edu/frontpage/}, and other fluxes are obtaining from references with wavelength center $\lambda_{\mathrm{center}}$. 
The Galactic extinction in the GALEX and optical bands were corrected based on \cite{fit99} and \cite{schlafly11} extinction map. 
$\spitzer$ and ISO spectrum are binned at the wavelength range shown in band width, and $\lambda_{\mathrm{center}}$ is the median wavelength at this wavelength range.
Band Width is $\lambda_{\mathrm{red}}$ - $\lambda_{\mathrm{blue}}$, where $\lambda_{\mathrm{blue}}$ is the blue side wavelength
with the half maximum transmission, and $\lambda_{\mathrm{red}}$ is the red side wavelength
with the half maximum transmission except $\spitzer$ and ISO data. 
PSF mean (large side) point spread function from references. References are (M): \cite{mar05}, \url{http://www.galex.caltech.edu/researcher/techdoc-ch1.html} and NED. (R): \cite{rod20} (Filter Profile Service: \url{http://svo2.cab.inta-csic.es/theory/fps/}) (F): \cite{flo04}. (O): \cite{ojh09}. (a): \cite{skr06}. (b): \cite{wri10}. (c): \cite{neu84} and \url{https://irsa.ipac.caltech.edu/IRASdocs/issa.exp.sup/ch4/C.html}. (d):  \cite{ona07}. (e): \cite{kaw07}. 
 (f): \cite{hou04} and \url{https://cassis.sirtf.com/atlas/query.shtml}. (g): \cite{kes96,bra08} and NED.
}
\end{deluxetable}

 \subsection{IR bands}\label{sec:h1821SED}

\subsubsection{2MASS Catalog}
The Two Micron All Sky Survey (2MASS\footnote{2MASS-IRSA/IPAC: \cite{2mass}};~\citealt{skr06})
covers 99.998\% of the celestial sphere in the near-infrared $J$ (1.25 $\um$), $H$ (1.65 $\um$), and $Ks$ (2.16 $\um$) bands. 
Bright source extractions have 1$\sigma$ photometric uncertainty of $<$~0.03 mag and astrometric accuracy in the order of 0.1~arcsec. We applied \verb|ph_qual=A|\footnote{2MASS documents: \url{https://irsa.ipac.caltech.edu/data/2MASS/docs/releases/allsky/doc/sec2_2a.html##ph_qual}}, which is equivalent to the S/N~$>10$, and all bands satisfy this criteria.

\subsubsection{ALLWISE Catalog}

The Wide-field Infrared Survey Explorer ($\wise$\footnote{$\wise$-IRSA/IPAC:~\cite{wise}};~\citealt{wri10}) mapped the entire sky in 3.4, 4.6, 12, and 22 $\um$ (W1, W2, W3, and W4) bands. 
The angular resolution is 6.1 arcsec, 6.4 arcsec, 6.5 arcsec, and 12.0 arcsec at 3.4, 4.6, 12, and 22 $\um$, respectively.
To obtain the reliable and clean flux without the contamination, \verb|ph_qual=A|\footnote{$\wise$ documents: \url{https://wise2.ipac.caltech.edu/docs/release/allwise/expsup/sec2_1a.html}} (S/N~$>10$) and \verb|ccflag=0| (unaffected by known artifacts) was applied, and all bands satisfy this criteria.

Due to the poor spatial resolution of the data in the W4 band, one should note the possibility that the emission from other sources may contaminate the flux measured in this band. In fact, we found one another optical source within 24 arcsec, which is the 2$\sigma$ range of the point spread function (PSF) of the W4 band. 
Although, this source shows \verb|ccflag|$\neq$0 (affected by known artifacts), we further checked the 2MASS sources within 24 arcsec, since the 2MASS gives the sharpest PSF among the IR photometries. We found only H1821 exists within 24 arcsec in the 2MASS catalog. 
This shows that the flux contamination of the W4 band from other sources, especially galaxies in the same cluster, is negligible.

\subsubsection{IRAS Catalogs}\label{sec:iras}
The Infrared Astronomical Satellite ($\iras$\footnote{$\iras$-IRSA/IPAC:~\cite{iras}};~\citealt{neu84}) mission performed an unbiased all-sky survey in the 12, 25, 60, and 100 $\um$ bands. 
The PSF of the instrument is about 5 arcmin\footnote{$\iras$ documents: \url{https://irsa.ipac.caltech.edu/IRASdocs/issa.exp.sup/ch4/C.html}}.
The typical position accuracy at 12 and 25 $\um$ is 7 and 35 arcsec in the scan and cross-scan direction, respectively. 
We found one counterpart whose PSF center of $\iras$ offsets from the optical one for 25~arcsec, which is consistent with the position accuracy at 25 $\um$.
In this paper we use data from the $\iras$/Point Source Catalog (PSC) except 12 $\um$ band, which does not have highest flux quality (\verb|fqual=3|). 

Because of the low spatial resolution in the
$\iras$ FIR bands, we also investigated the possible flux contamination from the nearby objects. 
There are 19 sources in the 2MASS catalog within 150~arcsec from the optical center of H1821 but their total fluxes contribute only 3.7\% of the W3 band flux of H1821. 
As a conservative estimate, we assumed a SED template for star-forming galaxies derived from \cite{rie09}: $f_{\mathrm{FIR,100 \um}}/f_{\mathrm{MIR,12 \um}} \sim 100$ and extrapolated the total FIR flux from the W3 band for all 19 objects.
The resulting flux contribution is 0.53 Jy at 100~$\mu$m, which is 24\% of the H1821 FIR (100 $\um$) flux. 


In addition to those point sources, there is a planetary nebula PN K 1-16 at an angular distance of 90~arcsec from H1821. 
The SED of the planetary nebula in the FIR can be described by black body spectra \citep[e.g.,][]{su07}. 
We found that the SED of PN K 1-16 is well fitted by a $\sim$~100~K black body spectrum, which is estimated from the flux ratio of the W2 and W3 bands. From this black body spectrum, the expected FIR flux contamination is only 3\% at 60 $\um$ and 10\% at 100 $\um$. These fluxes are subtracted from the $\iras$/PSC flux for the SED fitting of H1821 (see Section~\ref{sec:fitting}).  

Some might wonder how contamination in FIR fluxes affects the SED fitting result. 
We investigated it by changing photometric errors from the obtained value, 5\%, 10\% into 20\% plus 10\% systematic one (which is automatically implemented by \verb|CIGALE| code, see Appendix~\ref{app:err}). We found that it does not affect our results, which are summarized in Appendix~\ref{app:err}. Thus, we apply the obtained photometric errors plus 10\% systematic one for the SED fitting in this study.

\subsubsection{AKARI Point Source Catalogs}

 We compiled the MIR and FIR flux densities from the $\akari$ All-Sky Survey Point Source Catalogs. $\akari$ carries two instruments, the infrared camera (IRC\footnote{AKARI/IRC-IRSA/IPAC:~\cite{akari_irc}};~\citealt{ona07}) operating in the 2–26 $\um$ band (with two filters centered at 9 and 18 $\um$) and the Far-Infrared Surveyor (FIS\footnote{AKARI/FIS-IRSA/IPAC:~\cite{akari_fis}};~\citealt{kaw07}) operating in the 50–200 $\um$ band (with four filters centered at 65, 90, 140, and 160 $\um$). 
The spatial resolution of the FIS is $\simeq0.7$ arcmin for the 65 and 90~$\mu$m bands, and $\simeq1$ arcmin for the 140 and 160 $\mu$m bands.
In our study, we obtained only the fluxes with the highest quality with \verb|fqual=3|\footnote{See $\akari$ documents for more details \url{https://irsa.ipac.caltech.edu/data/AKARI/gator_docs/akari_fis_colDescriptions.html}}, leaving the fluxes only at 9, 18,
and 90 $\um$.

It is known that \textit{AKARI}/FIS 90~$\mu$m flux density differs by a factor of 2--3 with the $\iras$/PSC~65~$\mu$m and by a factor of 3--4 with the $\iras$/PSC~100~$\mu$m (see the release note of $\akari$/FIS\footnote{\url{https://irsa.ipac.caltech.edu/data/AKARI/documentation/AKARI-FIS_BSC_V1_RN.pdf}}),
which is also the case for H1821.
\cite{bra08} showed that \textit{Infrared Space Observatory} (ISO) fluxes are consistent with those of $\iras$/PSC at 100~$\mu$m band, and since we also added ISO data for our SED fitting, we only applied $\iras$/PSC data in this study and not using \textit{AKARI}/FIS 90~$\mu$m fluxes. 
We also show the result, which uses only $\akari$/FIS flux for FIR data in Appendix~\ref{app:sedfit}.

\begin{figure}
\begin{center}
\includegraphics[width=0.48\textwidth]{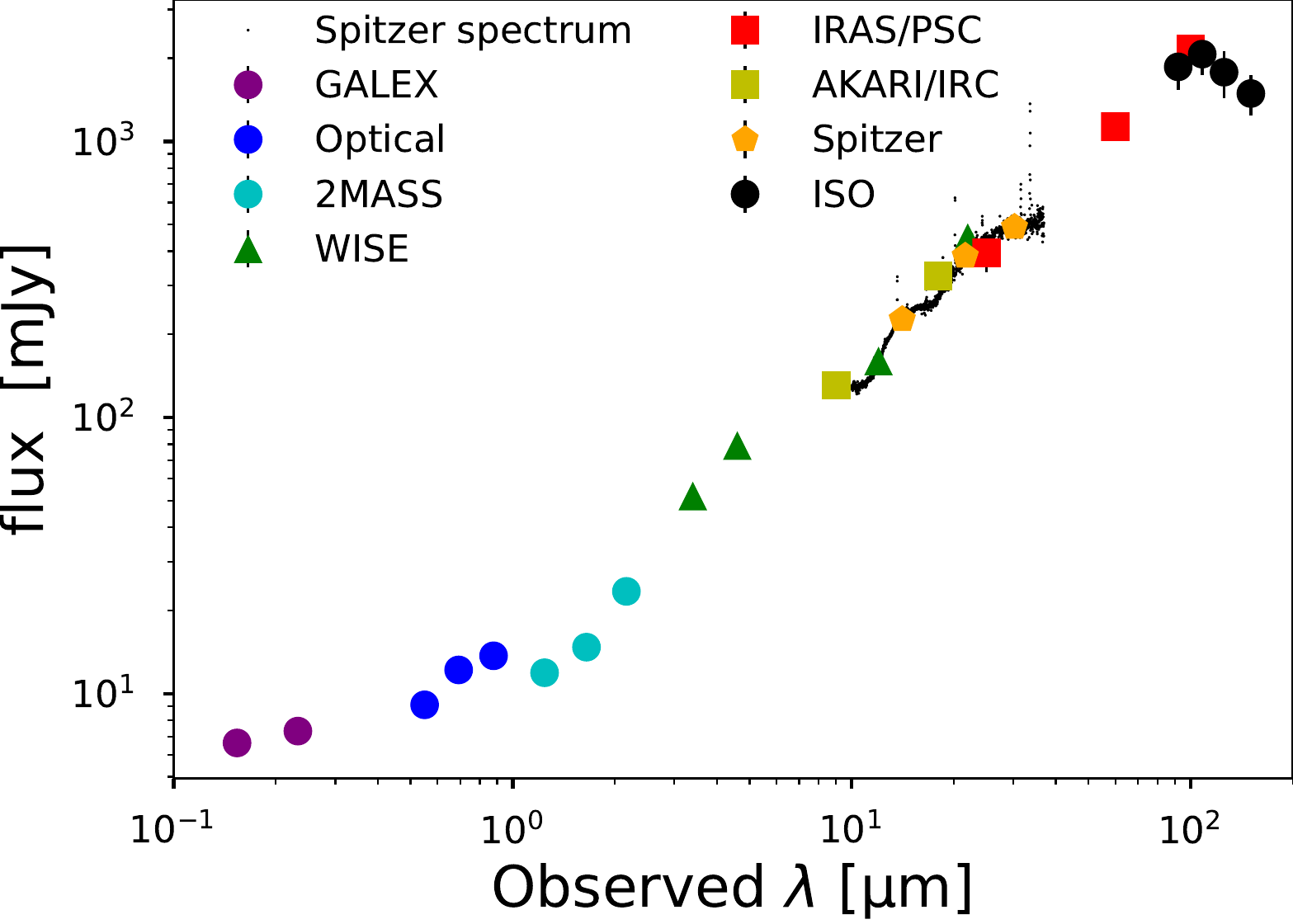}~
\caption{
UV, Optical and IR SED of H1821+643 with the $\spitzer$/IRS spectrum at 9--35~$\mu$m.
The different symbol represents the photometric data taken from different satellite/instruments.
}\label{fig:1821sed}
\end{center}
\end{figure}

\subsubsection{Spitzer and ISO IR spectra}\label{sec:spi}
We also obtained the spectra of $\spitzer$/Infrared Spectrograph \citep[IRS;][]{hou04} covering from 9 to 35~$\um$.
The Combined Atlas of Sources with $\spitzer$ IRS Spectra \citep[CASSIS;][]{leb11,leb15} provides the both of low-resolution (R$\sim$60-127) and high-resolution (R$\sim$600) ones for H1821. 
Since the SED fitting used in this study cannot be performed simultaneously with the spectroscopic data in \verb|CIGALE|, we binned the $\spitzer$ spectrum with good S/N (with S/N$>3$), into the three wavelength ranges at 11-18 $\um$, 18-26 $\um$, and 26-35 $\um$, to trace the silicate feature at 
$\sim20$~$\um$ \citep[e.g.,][]{hao07} and used those 3 bands for the SED fitting.

We also collected the FIR spectra of H1821 observed with the Long Wavelength Spectrometer (LWS; 43-195 $\um$) onboard the $\iso$ \citep{kes96}.
\cite{bra08} obtained the spectra covering 80--160~$\mu$m of H1821, and we made the $\iso$ spectrum into the four bins to trace the host galaxy dust emission. The more detailed analysis methods are described in Appendix~\ref{app:spi}.
We compiled the measured multi-wavelength properties in Table~\ref{tab:flux1} and the over-all SED is also compiled in Figure \ref{fig:1821sed}.

\subsection{UV, Optical and X-ray bands}
The UV photometries were obtained by Galaxy Evolution Explorer ($\galex$ \footnote{$\galex$ document:  \url{https://asd.gsfc.nasa.gov/archive/galex/instrument.html}}; \citealt{mar05}) 
in Far-UV (FUV, 135--175 nm) and Near-UV (NUV, 175--275 nm) bands and those information are summarized in the NASA/IPAC Extragalactic Database (NED\footnote{NED: \url{http://ned.ipac.caltech.edu/}}). 
There are two photometry data for each bands, and we utilize mean value for the SED fitting. The Galactic extinction coefficients in the GALEX bands, $A_\mathrm{FUV}=0.311$ mag and $A_\mathrm{NUV}=0.309$~mag,  were calculated with the \cite{fit99} reddening law with $R_V$ = 3.1 assuming a source spectrum of $f_\mathrm{\lambda} \propto \lambda^{-1.5}$. 

The optical V-band photometry was obtained by \cite{flo04} with the Hubble Space Telescope ($\hst$)/Wide Field and Planetary Camera 2 (centerd at 0.788 $\um$), and they converted the obtained magnitude to the standard Johnson V-band one. We did not show error in Table~\ref{tab:flux1}, because the error value was not reported. 
We also added the optical photometries obtained by \cite{ojh09} with the Naval Observatory Flagstaff Station (NOFS) 1.0 m Ritchey-Chretièn reflector in the Johnson R, and I bands. Optical photometries were corrected by using the \cite{schlafly11} extinction map ($A_V=0.118$, $A_R=0.093$ and $A_I=0.065$).

The absorption corrected X-ray fluxes at 2-10 keV and 14-195 keV bands, $F_\mathrm{2-10\ keV}$ and $F_\mathrm{14-195\ keV}$, are obtained from the BASS Survey X-ray spectral fitting analysis \citep{ric17} and summarized in Table~\ref{tab:flux_xray}. $F_\mathrm{2-10\ keV}$ and $F_\mathrm{14-195\ keV}$ are converted to
$f_{\nu} = 1.0 \times 10^{-3}$~mJy and $4.5 \times 10^{-5}$~mJy, respectively, assuming constant
$f_{\nu}$ in the energy range as described in the Equation (1) in \cite{yan20}.

\begin{deluxetable}{cccccccc}\label{tab:flux_xray}
\tabletypesize{\footnotesize}
\tablecolumns{4}
\tablewidth{0pt}
\tablecaption{X-ray properties of H1821+643 from \textit{Swift}/BAT \label{tab:flux_xray}}
  \tablehead{
        \colhead{$F_\mathrm{2-10\ keV}$}  
        & 
        \colhead{$F_\mathrm{14-195\ keV}$} &
        \colhead{colmun density} & \colhead{$\Gamma$} & \colhead{Ref.}
       \\
       \colhead{(erg~s$^{-1}$~cm$^{-2}$)} & \colhead{(erg~s$^{-1}$~cm$^{-2}$)} & \colhead{($\mathrm{cm^{-2}}$)} & \colhead{} 
     }
\startdata
 $1.97 \times 10^{-11}$ & $1.99 \times 10^{-11}$ &  $\leq 10^{20}$ &   2.22$^{\ +0.03}_{\ -0.02}$ &          a \\
\enddata
\tablecomments{Intrinsic X-ray fluxes, column density and spectral index ($\Gamma$) of H1821. (a): \cite{ric17}}
\end{deluxetable}

\section{SED fitting of H1821+643}\label{sec:fitting}

The near- (NIR) to mid-IR (MIR) bands provide information on the dust surrounding the central engine \citep{gan09,asm14,ich12,ich17}, while the far-IR (FIR) bands generally give the information on the cold dust in the host galaxies, except for several exceptions \citep[e.g.,][]{ros12,ich19}.
However, since star-formation from the host galaxy
sometimes contaminates the MIR emission, while the AGN
dust sometimes contaminate the FIR emission,
decomposition of the two components is important 
on the secured SFR and accretion rate estimation.

We apply the SED decomposition method to
obtain the AGN dust and host galaxy components
by using the \verb|CIGALE| SED fitting code, which builds the composite stellar/AGN spectrum through simple stellar populations with flexible star-formation histories (SFHs) and AGN radiation.
The fitting capabilities of \verb|CIGALE| \citep{boq19} were recently extended to X-ray energy bands \citep{yan20} to improve the characterisation of the AGN component.
In this study, we utilize the most up-to-date version called \verb|CIGALE| 2022.0 (hearafter \verb|CIGALE|) code \citep{yan22}.

For the AGN accretion disk and dust component,
\verb|CIGALE| utilizes SKIRTOR \citep{sta12,sta16} AGN model, which is one of the clumpy two-phase torus models based on 3D radiation-transfer code
and this model covers from the UV-to-far-IR emission of the AGN. 
Dust extinction and emission in the poles of type 1 AGNs \citep{bon12,lus12,sta19} are also considered. 
The X-ray emission is connected to the AGN emission at other wavelengths via the $\alpha_{\mathrm{ox}}$--$L_\mathrm{\nu}$(2500 {\AA}) relation of \cite{jus07}, 
where $L_\mathrm{\nu}$(2500 {\AA}) is the rest-frame 
2500 {\AA} luminosity, and $\alpha_{\mathrm{ox}}$ is the spectral slope between rest-frame 2500 {\AA} and X-ray (2 keV), defined as  $\alpha_\mathrm{OX}=0.3838 \times \log[(L_\mathrm{\nu}(\mathrm{2 keV}))$/$(L_\mathrm{\nu}$(2500 {\AA}))]. 
We refer the reader to \cite{yan20,yan22} for a full description of \verb|CIGALE|. The modules and grid of parameters we used in our analysis are summarized in Table \ref{tab:para}, and other parameters not mentioned in this study are same as \verb|CIGALE| default values. 
These grid of models are fitted to the observational data in \verb|CIGALE|, and \verb|CIGALE| estimates reduced chi square for each parameter.
The physical values are based on the likelihood-weighted means of the parameters obtained by the \verb|CIGALE| fittings, and the errors are based on the standard derivations of the obtained parameters.
Here, we describe the main steps to build the models and fit the SED covering from the X-ray to FIR bands.

\begin{deluxetable*}{lcc}
\tablecaption{Models and the values for free parameters used by CIGALE for the SED fitting of H1821+643 and the values selected as the best model.}
\tablewidth{0pt}
 \label{tab:para}
\tablehead{
\colhead{Model/Palameter} 
 & \colhead{Values} & \colhead{select}
}
\decimalcolnumbers
\startdata
   \multicolumn{2}{c}{Star-formation history: double-exponentially decreasing ($\tau$-decay) model}\\
   e-folding time of the main stellar population (Myr)& 500, 1000, 2000, 3000 & 500\\
    Age of the main stellar population (Myr)& 4000, 6000, 8000, 10000& 10000\\
      e-folding time of the late starburst population  (Myr)& 25, 50, 75, 100 & 100\\
    Age of the late burst  (Myr)& 5, 10, 20, 50, 100 & 5\\
    Mass fraction of the late burst population & 0.01, 0.03, 0.1& 0.1\\
     \hline 
     \multicolumn{2}{c}{Stellar population synthesis model}\\
    Simple stellar population & \cite{bru03}&\\
    Initial mass function & \cite{sal55}&\\
    Metallicity ($Z$) & 0.4 $Z_{\odot}$, $Z_{\odot}$ & 0.4~$Z_{\odot}$\\
     \hline 
    \multicolumn{2}{c}{Galactic dust attenuation: \cite{cal00}}\\
    Colour excess of stellar continuum light for young stars E(B−V) & 0.1-0.6 (step 0.1) & 0.2\\
    \hline 
    \multicolumn{2}{c}{Galactic dust emission: \cite{dal14}} \\
    $\alpha$ slope in $dM_{dust} \propto U^{-\alpha} dU$  &  1.5, 1.75, 2.0 & 1.75\\
    (IR power-law slope)\\
     \hline 
    \multicolumn{2}{c}{AGN (UV-to-IR): SKIRTOR \citep{sta12,sta16}}\\
 Viewing angle ($\theta$)  & $30^{\circ}, 40^{\circ}$ & $40^{\circ}$\\
 (face on: $\theta$ = $0^{\circ}$, edge on: $\theta$ = $90^{\circ}$)\\
 Deviation from the default UV/optical slope ($\delta_\mathrm{AGN}$) & -0.6 to 0 (step 0.1) & -0.2 \\
 a AGN fraction in total IR luminosity & 0.4, 0.5, 0.6, 0.7& 0.6\\
 Extinction law of polar dust & SMC &\\
 E(B − V) of polar dust & 0.0, 0.03, 0.1 & 0.03\\
Temperature of polar dust (K) & 100, 150, 200 & 150\\
  Emissivity of polar dust & 1.6 &\\
     \hline 
  \multicolumn{2}{c}{AGN X-ray}\\
 AGN photon index ($\Gamma$) & 2.2 & \\
 X-ray energy cut-off ($E_\mathrm{cut}$) & 330 keV \\
 $\alpha_{\mathrm{ox}}$ &  -1.4 to -1.2 (step 0.05)& -1.25\\
 Maximum deviation from the $\alpha_{\mathrm{ox}}$--$L_\mathrm{\nu}$(2500 {\AA}) relation & 0.5 &\\
\enddata
  \thispagestyle{empty}
\label{tab:test11}
\end{deluxetable*}

\subsection{Host Galaxy Component}\label{sec:sfh}

The host galaxy component can be characterized by the combination of the assumed SFH and the initial mass function (IMF) of the stellar emission.
On the SFH, we applied the delayed SFH with optional exponential burst to characterize the SED of ULIRGs with the experience of recent starbursts \citep[e.g.,][]{rie09,cie15, boq19}.

The stellar emission is modelled using the \cite{bru03}, and we adopted the Salpeter IMF \citep[][mass range 0.1--100 $\msun$]{sal55} to keep the consistency with the SFR estimation using \cite{ken98} equation (see Section~\ref{sec:SFR}). 
The metallicity of the gas ($Z$) at the $\sim$~30~kpc from the H1821 center was found to be $Z \sim 0.4\ Z_{\odot} = 0.008$ in the observations \citep{rus10}, where $Z_{\odot}$ represents the solar metallicity. 
In addition to this value, we set the solar metallicity $Z_{\odot}$ to account for the contamination/pollution caused by star-formation in H1821.
The stellar emission was attenuated with the \cite{cal00} attenuation law. 
The IR SED of the dust heated by stars was implemented with the \cite{dal14} template. 
We added the nebular-emission components to the SED by using ‘nebular-emission’ module.

\subsection{AGN Component}\label{sec:agn}

We used the built-in SKIRTOR AGN model \citep{sta12,sta16} to fit the AGN emission from the UV to FIR bands. The SKIRTOR AGN model covers the most recently updated polar AGN dust emission which has been recently suggested from the MIR interferometric observations \citep[e.g.,][]{bur13,hon12,hon13,hon19}.
Although the dust grain size distribution is not known yet for the polar dust emission \citep[but see][for the detailed discussion of the dust size distribution]{lyu18,taz20a,taz20b},
we here adopted the Small Magellanic Cloud (SMC) extinction curve \citep{pre84} for the polar dust extinction curve, since it is preferred from AGN observations \citep[e.g.,][]{hop04,sal09,bon12} and it is also the default manner of \verb|CIGALE|.
We set polar dust temperature at 100, 150, 200~K for tracing the MIR AGN dust emission peak with emissivity index of 1.6 \citep{yan20}. 

We adopted $\alpha_{\mathrm{ox}}=-1.4$ to $-1.2$ for the fitting with a step resolution of $\Delta \alpha_\mathrm{ox}=0.05$,
whose range was taken from the obtained observational lower-limit value of $\alpha_{\mathrm{ox}}=-1.4$ for the case of H1821. 
Note that this value deviates roughly 3$\sigma$ from the scatter of the known $\alpha_{\mathrm{ox}}$--$L_\mathrm{\nu}$(2500 {\AA}) relation, which suggests $\alpha_{\mathrm{ox}}=-1.7$.
This indicates that 
H1821 has small bolometric correction with high X-ray luminosity, or H1821 might be slightly obscured in UV wavelength, as suggested from the fact that H1821 is classified as Seyfert 1.2 galaxy. 
Therefore, the viewing angle is set to $30^{\circ}, 40^{\circ}$, a relatively edge-on view type-1 AGN or the reddening towards the broad line region \citep{ver06}. 
In addition, in order to describe the blue AGN color, we allow $\delta_\mathrm{AGN}=-0.6$ to $0$ (with the step of 0.1), where $\delta_\mathrm{AGN}$ is the deviation from the slope of SKIRTOR model at UV-optical continuum.

\subsection{AGN X-ray Emission}\label{sec:gal}

\verb|CIGALE| accepts the X-ray flux information to estimate the original accretion disk emission assuming the $\alpha_\mathrm{OX}$ as discussed above.
AGN X-ray radiation typically has a energy cut oﬀ ($E_\mathrm{cut}$) in the power law, i.e.
$f_{\nu} \propto E^{-\Gamma + 1}\ \mathrm{exp}(-E/E_\mathrm{cut})$. 
\cite{ric17} obtained $\Gamma$ = $2.22^{+0.03}_{-0.02}$ and lower limit of $E_{\mathrm{cut}}\geq$~130 keV for H1821.
Thus, we set $\Gamma$ = $2.2$ and $E_{\mathrm{cut}}$=330~keV, whose 
$E_{\mathrm{cut}}$ values are obtained from the mean values of unobscured AGNs in the local universe \citep{ric17}.


\section{result}\label{sec:result}

\subsection{SED Fitting Result}\label{sec:sedfit}

\begin{figure}
\begin{center}
\includegraphics[width=0.48\textwidth]{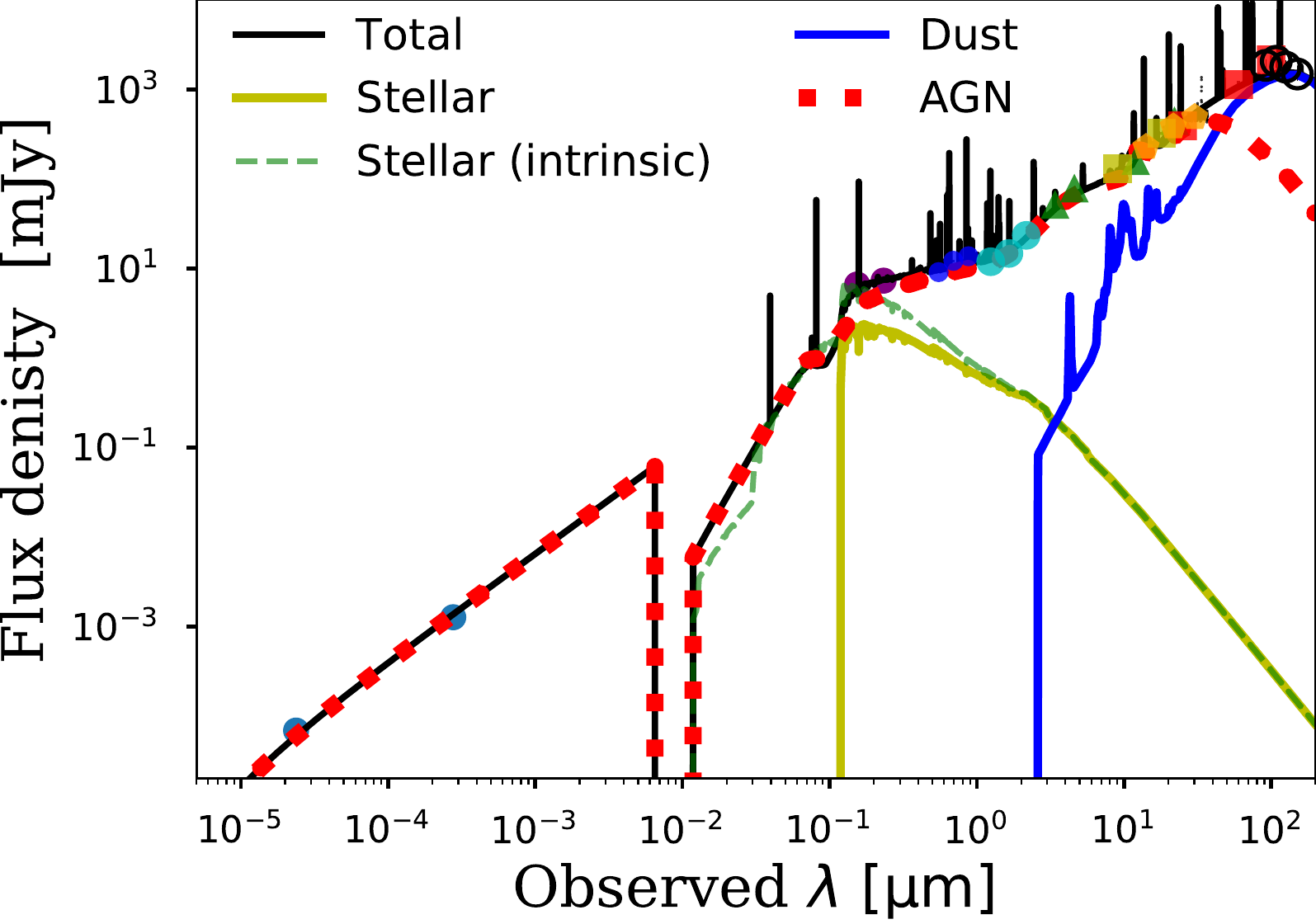}~
\caption{SED fitting result of H1821. The filled points are the photometric data and the symbols are same as in Figure~\ref{fig:1821sed}.
The red dotted line represents the AGN direct and dust emission. The blue solid line represents the dust emission from the host galaxy. 
The yellow line represents the stellar emission from the host galaxy and 
the green dashed 
line represents the intrinsic stellar emission.
The black solid line is the combined one of dust, stellar, nebular, and AGN.
}\label{fig:sedre}
\end{center}
\end{figure}

 Figure~\ref{fig:sedre} shows the SED fitting result for the best model of H1821 by \verb|CIGALE|, with the decomposed AGN (red dotted line), stellar (yellow solid line), and host dust (blue solid line) components. Here the AGN component is the sum of AGN disk and AGN torus emission. We also show intrinsic stellar component as green dashed line. 
 Physical quantities obtained by SED fitting are shown in Table~\ref{tab:physical_value}. 
 The total IR luminosity reaches $\log ( L_\mathrm{IR,tot}/\lsun)
=13.15 \pm 0.06$, 
 indicating H1821 is HyLIRG, which is consistent with the previous studies \citep{row91,san96}.

\subsubsection{SFR and stellar mass}\label{sec:SFR}

The obtained dust luminosity of the host galaxy after the removal of the AGN dust component, reaches 
$\log ( L_\mathrm{host,IR}/\mathrm{erg}~\mathrm{s}^{-1})
={46.35 \pm 0.04}$.
The SFR is estimated from $L_\mathrm{host,IR}$ based on the conversion equation by \cite{ken98}, with $\log (\mathrm{SFR}/\msun~\mathrm{yr}^{-1}) = \log ( L_\mathrm{host,IR}/\mathrm{erg}~\mathrm{s}^{-1})-43.34$,
and the obtained SFR is 
$\log ( \mathrm{SFR}/\msun~\mathrm{yr}^{-1}) = 3.01 \pm 0.04$,
indicating that H1821 is in an extreme starburst phase and is in an extreme environment achieving the highest SFR in the local universe at $z<0.4$.
The obtained SFR is larger by than that of \cite{rui13}
of SFR $= 300^{+300}_{-200} \ \msun$~yr$^{-1}$ estimated from the PAH~7.7~$\mu$m emissions in the $\spitzer$/IRS MIR spectra.
Note that the PAH based SFR is difficult for the quasar such as H1821 since the PAH emission is overwhelmed by the underlying AGN dust continuum
in the case of H1821, which makes the estimation of the SFR difficult
and it is why the uncertainty of the PAH-based SFR is significant.
On the other hand, the obtained SFR in this study is consistent with that of IR SED fitting studies by \cite{far02} of SFR$=1100\pm 200 \ \msun$~yr$^{-1}$.

\begin{figure}
\begin{center}
\includegraphics[width=0.48\textwidth]{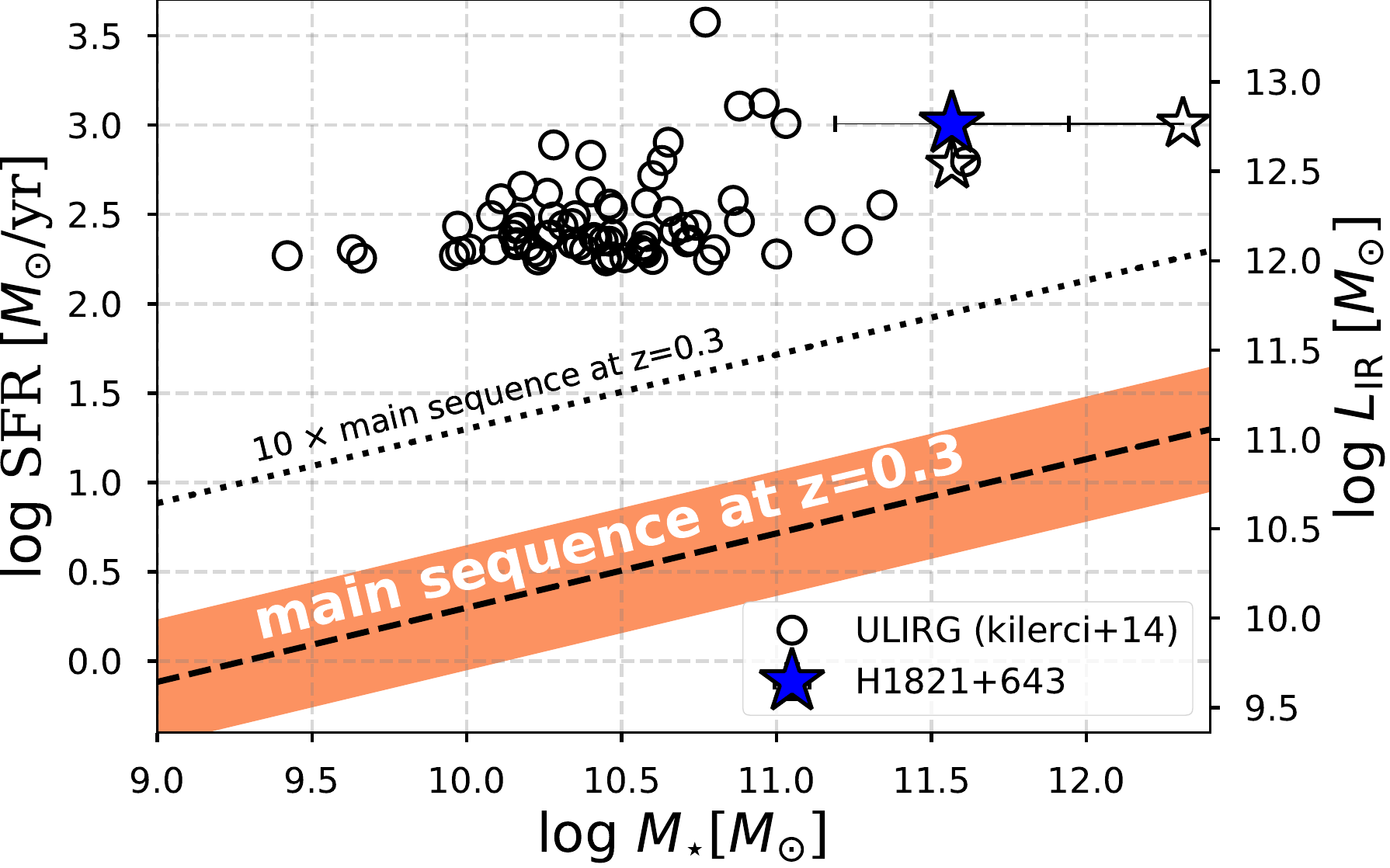}~
\caption{The relation between SFR and
stellar-mass ($\mstar$) of H1821, plotted as blue star, and the 
star-formation main sequence at $z \sim 0.3$ \citep{pea18}. 
The lowest SFR of H1821 is estimated using $\akari$ data for FIR photometry, and the highest stellar mass is the value obtained from $\hst$ PSF subtraction method.  
The shaded area is the expected main-sequence (MS) region at $z=0.3$ with the scatter of 0.35 dex. 
Empty circles are the local ULIRG and one HyLIRG obtained from \cite{kil14} with redshift 0.050 $<z<$ 0.487 (median is 0.181). The dotted line represents the 10 times above the MS at $z=0.3$.}
\label{fig:SF_MS}
\end{center}
\end{figure}

Figure~\ref{fig:SF_MS} shows the location of H1821 in the SFR and $\mstar$ plane, where both values are obtained from the \verb|CIGALE| fitting. 
The SFR of H1821 estimated by $\akari$ data is also shown with the open star at $\log (\mathrm{SFR}/\msun~\mathrm{yr}^{-1})=2.78\pm 0.02$.
It should be noted that the NIR SED is dominated by AGN emission and thus the estimation of $\log (\mstar/\msun)= 11.57 \pm 0.38$ 
has a large uncertainty.
Actually, the obtained $\mstar$ is 2 to 10 times lower than the value of $\mstar=2.05 \times 10^{12}\ \msun$ estimated by $\hst$-PSF subtracted method \citep{flo04}, which is shown with a open star in Figure~\ref{fig:SF_MS}.
This difference would originate from the
difference of the assumed mass-to-light ratio,
not because of the difference of the obtained stellar emission contribution.
For example, our SED fitting indicates that the stellar component contributes 4-9\% and 14-16\% of the emission at $H$- and $V$ band,
which are almost consistent with the $\hst$--PSF subtracted method with 10\% and 7.5\% at $H$- and $V$ band, respectively.
On the other hand, \cite{flo04} assumed mass-to-light ratio of an early-type galaxy with a stellar age of $\sim10$~Gyr, while the \verb|CIGALE| 
fitting shows a mass-weighted stellar age of $\sim4$~Gyr.
This produces the difference of the mass-to-light ratio
of a factor of $\sim3$, which is almost consistent with
the difference of the stellar-mass between the two methods.

It is known that most star-forming galaxies follow the main-sequence (MS), and sources above the MS are called starburst galaxies \citep[e.g.,][]{elb11,pea18}.
Although $\mstar$ and SFR has a large uncertainty, this extremely high SFR indicates that H1821 is well above the expected SF main-sequence at $z=0.3$, which is shown as orange shaded area with 1$\sigma$ scatter in Figure~\ref{fig:SF_MS} obtained from \cite{pea18}.

In the local universe, one observationally known population in this starburst locus is
ultra/hyperluminous infrared galaxies (U/HyLIRGs) whose IR luminosity reaches $L_\mathrm{IR} \geq 10^{12}\ L_\odot$ and  $L_\mathrm{IR} \geq 10^{13}\ L_\odot$, respectively. This extreme IR luminosity corresponds to SFR~$\gtrsim 170-1700\ \msun$~yr$^{-1}$ assuming the empirically known $L_\mathrm{IR}$--SFR relation \citep{ken98}.
Figure~\ref{fig:SF_MS} shows that 
U/HyLIRGs in $0.1<z<0.5$ reside in above the MS by a factor of roughly 10--100, at $z\sim0.3$ \citep{kil14}.
Copmared to U/HyLIRGs, H1821 nicely locates the top-end of the SFR range and heavier end of the
stellar-mass. This indicates that H1821 is in extremely efficient star-formation environment, where only a small fraction of local ULIRGs can achieve.
Thus, either the method used by this study or by the $\hst$-PSF decomposition method, it is safe to conclude that H1821 is in starburst mode, indicating the indirect evidence for the availability of cold gas supply fueling star-formation.

\begin{deluxetable*}{cccccccc}
\tabletypesize{\footnotesize}
\tablecolumns{6}
\tablewidth{0pt}
\tablecaption{Result of SED fitting \label{tab:physical_value}}
  \tablehead{
      \colhead{Reduced chi square} &\colhead{$\log L_{\mathrm{AGN,IR}}$} &\colhead{$\log L_{\mathrm{AGN,bol}}$} 
      &\colhead{$\log \mdotbh$}
      & \colhead{$\log L_{\mathrm{host,IR}}$} &\colhead{$\log \mstar$} & 
\colhead{$\log$~SFR} 
      \\
       \colhead{} & \colhead{(erg~s$^{-1}$)} & \colhead{(erg~s$^{-1}$)}
       & \colhead{($\msun$~yr$^{-1}$)} &\colhead{(erg~s$^{-1}$)} & \colhead{($\msun$)} & 
       \colhead{($\msun$~yr$^{-1}$)}
     }
\startdata
 1.0 &  46.51 $\pm$ 0.04 &  46.96 $\pm$ 0.05 &  1.20 $\pm$ 0.05 &  46.35 $\pm$ 0.04 &  11.57 $\pm$ 0.38 &  3.01 $\pm$ 0.04 \\
\enddata
\tablecomments{Physical quantities obtained by SED fitting using $\iras$/PSC and $\iso$ fluxes for FIR data with photometric uncertainty.
}
\end{deluxetable*}

\subsubsection{AGN luminosity and BH accretion rate}\label{sec:result_AGN}

The \verb|CIGALE| fitting also gives the AGN dust luminosity with $\log ( L_\mathrm{AGN,dust}/\mathrm{erg}~\mathrm{s}^{-1})
={46.51 \pm 0.04}$ 
and the bolometric AGN luminosity of $\log ( L_\mathrm{AGN,bol}/\mathrm{erg}~\mathrm{s}^{-1}) = 46.96 \pm 0.05$ 
, which is almost comparable with the typical luminosities of the bright SDSS quasars at $z\sim1$--$2$ \citep{she11}. 
The obtained Eddington ratio is $\eddington=0.18$, which is smaller than the value estimated from the X-ray based value of $\eddington=0.59$ shown in Section~\ref{sec:intro}.
Still, it is considered to be in a radiatively efficient state \citep[e.g.,][]{ina20}.
Hereafter, we use $\eddington=0.18$ as a fiducial value in this study.
In this Eddington ratio value, where the standard disk state is considered to be realized, 
the energy output can be dominated mainly by radiation not by jets. We further discuss how AGN jets heating on the intra-cluster medium (ICM) affects our results in Section~\ref{sec:5.1_heating}.

Interestingly, as discussed above, this bolometric AGN luminosity indicates a small bolometric correction $k_\mathrm{2-10 keV}$ of 
$L_\mathrm{AGN,bol} = 17  L_\mathrm{2-10keV}$ (see Section~\ref{sec:intro}) with higher X-ray to MIR ratio, $L_\mathrm{2-10 keV}/\nu L_\nu (6 \um) = 0.5$.
This is an opposite trend to the high-luminosity AGN such as SDSS quasars with the same MIR luminosity range, which usually show a saturation of the X-ray luminosity compared to the bolometric one \citep{ric17a,tob19,ich22}. 
The trend of the high X-ray emission is also seen from flatter 
$\alpha_\mathrm{OX}$ with $\alpha_\mathrm{OX}=-1.25$ as suggested in Section~\ref{sec:agn}, compared to the typical quasars with the same UV luminosity range $L_\mathrm{\nu}$(2500 {\AA})~$\sim 2 \times 10^{31}$~erg~s$^{-1}$~Hz$^{-1}$ \citep{jus07,vas07}. 
This discrepancy may be related to the very massive $\mbh$ in H1821 with $\mbh = 3.9 \times 10^9 \ \msun$ and thus it has a relatively smaller Eddington ratio ($\eddington=0.18$) compared to other high-luminosity AGN system, with $\eddington\simeq 0.3$. 
In fact, the bolometric correction of H1821 is consistent with the values estimated from their Eddington ratio \citep{vas07,tob19}.

The SMBH in H1821 is already matured, but it still has a rapid gas accretion rate to the BH. Current SMBH mass of $\mbh = 3.9 \times 10^9 \ \msun$ is already close to the largest mass BHs that have been found only in the center of galaxy clusters or groups in the local universe \citep[see][]{mcc11,tho16}. 
This suggests that the large amount of gas in the cluster (center) makes it a unique location for the growth of massive BHs that cannot be produced in other environments.

We also estimate the gas consumption rate at BH scale, which can be derived from the BH accretion rate $\mdotbh$:
  \begin{equation}
\label{equ:mdotbh}
      \mdotbh = \frac{\lbol}{\eta c^2}, 
\end{equation}
 where $\eta$ is the radiation efficiency, and we adopt a canonical value of $\eta=0.1$ \citep{sol82}.
The obtained accretion rate is 
$\log (\mdotbh/\msun~\mathrm{yr}^{-1}) = 1.20 \pm 0.05$, 
indicating that large amounts of gas are available even in much smaller BH sphere of influence with $\ll 100$~pc. 
From the obtained BH accretion rate,
the gas consumption ratio reaches $\mdotbh/\mathrm{SFR}\sim~0.02$,
which is almost comparable with the one obtained
 by the BCGs of $\mdotbh/\mathrm{SFR}\sim0.04$ \citep{mcd21},  even though the absolute values of $\mdotbh$ and SFR is considerably smaller in the BCGs. 
 In addition, the obtained gas consumption ratio
 is also similar value with some most massive
BCGs showing a high BH mass to bulge mass ratio reaching up to 0.01--0.2 \citep[e.g., NGC 4486B and NGC 1277;][]{kor13}.
On the other hand, the gas consumption ratio is larger by a factor of 4--7 than the local BH mass to bulge mass ratio \citep[$\sim$ 0.005][their Section 6.6.1]{kor13} and the ratio found in the local Seyfert galaxies (e.g., 0.003 in \citealt{dia12}), which studied SFR within the bulge scale. 
Combined the results above,
H1821 has a higher gas consumption ratio compared to other galaxies and AGN in the field. That is,
the cluster environment such as H1821 might produce a slightly higher BH to stellar-mass ratio in the central galaxy
and if H1821 keeps the current gas consumption ratio, H1821 is likely to join as one of the BCGs which has the most highest BH mass to bulge mass ratio.

Given the result that the
total cold gas consumption rate reaches
$\dot{M}_\mathrm{con} = \mdotbh + \mathrm{SFR} = 10^3$~$\msun$~yr$^{-1}$, H1821 is in an cold gas rich environment both in the scale of the host galaxy ($\sim10$~kpc) and the black hole scale ($\ll 100$~pc).


\subsection{Gas Supply to Generate/Keep Active Star-Formation}\label{sec:gas}

Section~\ref{sec:sedfit} summarizes that H1821 is in a gas-rich environment and H1821 drastically consumes the supplied gas with $\sim 10^3\ \msun$~yr$^{-1}$.
This raises the question of how such cooling gas is provided in H1821 and how long such drastic gas consumption has lasted and will last.
We investigate here the ratio of gas consumption rate over the gas cooling rate of the intra-cluster medium (ICM)  $\dot{M}_\mathrm{con}/\dot{M}_\mathrm{cool}$, which is an indicator of how efficiently gas is consumed once it is provided to the galaxy by the cooling of the gas in the cluster center through emitting X-ray radiation dominated by Bremsstrahlung and metal lines \citep[e.g.,][]{fab94}. We note that the cooling rate of the cluster gas is estimated traditionally within cooling radius $r_\mathrm{cool}$: the cooling time becomes 7.7~Gyr, which is the time since the $z=1$ (Section~\ref{sec:intro}, \citealt{hud10}). Thus, this cooling rate is the long-time averaged supplying rate of the cluster gas.

The gas cooling rate of H1821 is obtained as $\mdotcool= (3.0 \pm 1.0) \times 10^2\ \msun$~yr$^{-1}$ within the cooling radius of 90~kpc by \cite{rus10}, which conducted high spatial resolution X-ray observation by \textit{Chandra} satellite and data analysis using \textit{XSPEC mkcflow}. 
\cite{mcd18} also did a Chandra data analysis and obtained $\mdotcool= (4.7 \pm 0.5) \times 10^2\ \msun$~yr$^{-1}$ which is comparable to the \cite{rus10} analysis, although there are few differences between \cite{mcd18} and \cite{rus10} assumptions. 
Based on these values, the cooling rate of $\sim 300-500\ \msun$~yr$^{-1}$ is 2-3 times smaller than the gas consumption rate of $\dot{M}_\mathrm{con}\sim 10^3\ \msun$~yr$^{-1}$.
This high gas consumption efficiency of 
$\dot{M}_\mathrm{con}/\dot{M}_\mathrm{cool} \gtrsim 1$ indicates that H1821 is in a rapid gas consumption phase by slightly exceeding the available all gas supply.

\begin{figure}
  \begin{center}
    \includegraphics[width=0.48\textwidth]{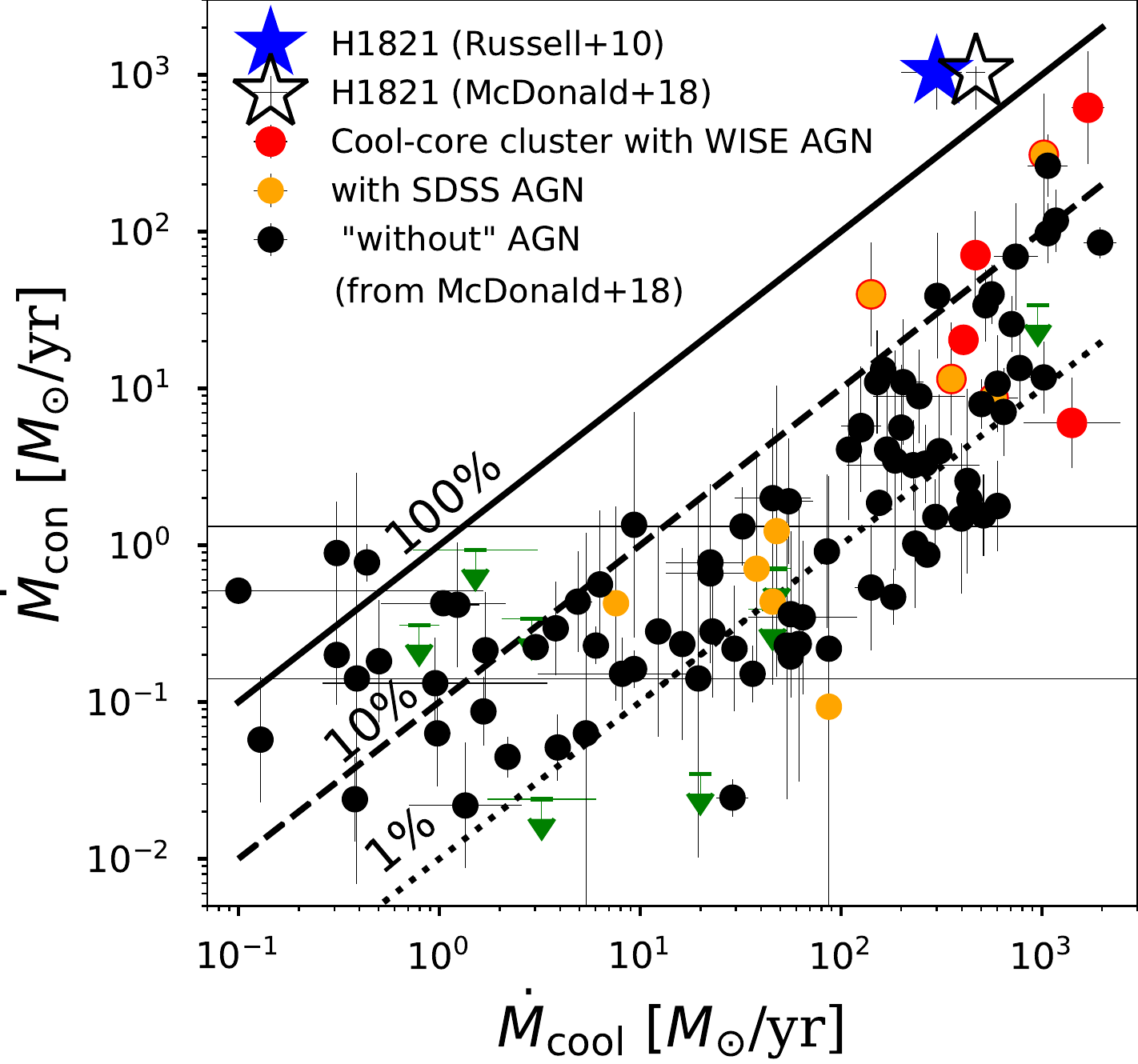}\\
    \caption{Gas consumption rate of the central galaxy as a function of the cooling rate of ICM, for 104 galaxies, groups, and clusters from \cite{mcd18} for comparison with H1821. The cooling rate of
H1821 is obtained from \citet[][blue filled star]{rus10} and \citet[][black open star]{mcd18}, respectively. The gas consumption rate of $\dot{M}_\mathrm{con}=\mdotbh + \mathrm{SFR}$ is obtained from our SED fitting results for H1821. For the other sources, we treat $\dot{M}_\mathrm{con}=\mathrm{SFR}$  \citep{mcd18} since $\dot{M}_\mathrm{con}$ is dominated by $\mathrm{SFR}$ and the availability of $\mdotbh$ is not complete.
    Red circles show systems which were found AGN activity based on $\wise$ color, and orange circles indicate systems defined as AGN by SDSS spectra. Note that some systems do not have SDSS spectra, and BCG ``without AGN'' mean systems, which do not show red $\wise$ color and not defined as AGN based on SDSS spectra or not have SDSS spectra. Green arrows show upper limits of gas consumption rate by AGN contamination and detection limit. Typical gas consumption rate is 1-10\%, but H1821 consumes $\gtrsim$~100\% gas.}
    \label{fig:compare}
  \end{center}
\end{figure} 

We also compare the ratio of gas consumption rate over the gas cooling rate $\dot{M}_\mathrm{con}/\dot{M}_\mathrm{cool}$ with other 104 
galaxies in the \cite{mcd18} sample, as shown in Figure \ref{fig:compare}. 
They assembled galaxies, groups, and clusters from the literature,
which span a wide range in mass, cooling rate $\dot{M}_\mathrm{cool}$, redshift ($0 < z < 1$) and AGN activity \citep{rus13, cav09, fog17, fra14}.
 The sample has a range of $\dot{M}_\mathrm{cool}=$~0.1--2000~$\msun$~yr$^{-1}$ and various AGN activity with $\eddington \sim$~$10^{-6}$ to 1. 
 The sample is divided into subgroups based on SDSS spectra and/or $\wise$ IR colors, such as SDSS~quasar which shows a broad emission line (orange filled circle), SDSS~(type-2) AGN based on the emission line ratio diagram \citep[so called BPT diagram; orange filled circle; ][]{bal81}, and $\wise$ selected AGN (red filled circle).
Either if the sources do not fulfill the AGN criteria above, or if sources do not have SDSS spectra, they are labeled as BCG ``without AGN'' (black filled circle),  which means that bright AGN might contaminate in this subgroup. 
 
As shown in Figure~\ref{fig:compare}, the gas consumption efficiency $\dot{M}_{\mathrm{con}}/\dot{M}_{\mathrm{cool}}$
of typical cool core cluster 
has a range of 
$\dot{M}_\mathrm{con}/\dot{M}_\mathrm{cool} \sim 10^{-2}$--$10^{-1}$. 
In contrast, Figure~\ref{fig:compare} illustrates that 
H1821 is located in the unique parameter range
$\dot{M}_{\mathrm{con}}/\dot{M}_{\mathrm{cool}} \gtrsim 1$. 
The actual gas mass rate that is completely cooling is in general smaller than that of $\dot{M}_{\mathrm{cool}}$, with $\dot{M}_{\mathrm{cool,true}}\lesssim \dot{M}_{\mathrm{cool}}/10$ \citep{pet03, rus10},  
thus $\dot{M}_{\mathrm{con}}/\dot{M}_{\mathrm{cool,true}} \gg 1$ for the case of H1821.
This suggests that H1821 is in a special
environment, reaching the highest gas consumption
efficiency, which produces the extreme SFR well above the SF main sequence and even above/comparable to the starburst galaxies at the same redshift.

\section{Discussion}\label{sec:dis}
\subsection{Energy Balance between Cooling and Heating in the Central 100~kpc of H1821}\label{sec:5.1_heating}
Most central galaxies in galaxy clusters are already quenched in the local universe \citep{pet03} and energy injection of AGN jets may play an important role to maintain the cooling and heating balance of the gas in the galaxy clusters \citep[e.g.,][]{fab12}. 
As discussed in Section~\ref{sec:gas}, 
H1821 is a unique object with 
its high gas consumption efficiency $\dot{M}_{\mathrm{gas}}/\dot{M}_{\mathrm{cool}}$ $\gtrsim$ 1, while other central galaxies in galaxy clusters could use only a few percent of the gas.
The main origins of such small consumption efficiency for most of the galaxy clusters are considered to be due to heating from AGN jets.
Actually, there are observational suggestions that most galaxy clusters show prominent jets with association of X-ray cavities, hampers the gas from the cooling \citep[e.g.,][]{bir04, dun06}. 
A prominent radio jet and the $\sim10$~kpc scale X-ray cavity is also reported for H1821 \citep{blu96, blu01, rus10}, suggesting that the jet might further suppress the gas cooling also for H1821.

Although the (time-averaged) gas cooling rate is still not enough to sustain the current rapid gas consumption in H1821, it is important to investigate whether the cooling rate is further suppressed by AGN jets also for H1821.
We evaluate the energy balance between the cooling and heating (by AGN jet) in the H1821 system
with $\sim100$~kpc scale.
\cite{rus10} estimated the heating of the AGN jet as $W_{\mathrm{heat, jet}} = $
(1--2)$\times 10^{44}~\mathrm{erg}~\mathrm{s}^{-1}$, while the cooling rate is $W_{\mathrm{cool}} \sim 9 \times 10^{44}$~erg~s$^{-1}$ within the cooling radius of 90~kpc.
Thus, the cooling is larger than the AGN jet heating $W_{\mathrm{cool, ff}} / W_{\mathrm{heat, jet}} = $~(4.5--9).
In addition, thanks to the high accretion rate into the BH
with $\eddington\sim 0.2$, the standard disk state is likely realized in H1821 and this disk state releases energy mainly as radiation rather than as jet \citep[e.g.,][]{ho08,lis19,ina20}.
Those results indicate that most of the gas in the H1821 system can be cooled without the interruption of the heating by AGN jets.

\subsection{FIR Fine Structure Line: Cooling Path to Star-Formation}\label{sec:fine}

The surroundings of H1821 is an ideal place for the gas to cool down 
thanks to the efficient cooling through X-ray radiation without suppression, on average, by AGN jets. We here discuss extended FIR fine structure emission lines, which can
be a prominent coolant in this system.
Although the Bremsstrahlung is a dominant cooling process for the gas down to $T\sim10^{7}$~K
\citep[e.g.,][]{sut93}, 
star-formation requires much cooler gas below 100~K \citep[e.g.,][]{leo08,schruba11},
which still has a huge temperature gap with more than five orders of magnitude.
Considering that the cooling time of Bremsstrahlung of cluster scale is order of $\sim$~Gyr,
there should have 
efficient cooling process achieving from $10^7$~K to $100$~K
within $<$~Gyr to realize the prompt star-formation as seen in H1821.

\cite{sut93} calculated the net cooling function of a plasma, including 
both heating and cooling processes in the temperature range of $10^{4}$ K to $10^{8}$ K. 
According to their calculations, the net cooling function rapidly increases from $10^{7}$~K because of the emission process of the Fe group lines at high temperature ($T\sim 10^{6}$ K) and lighter atoms such as C, O, and Ne in the lower temperature range ($10^{4.5}$ K to $10^{6}$ K). 
In the temperature range around $10^{4}$ K, Ly$\alpha$ determines the cooling function.
Thus, once the gas cools down with $\sim10^7$~K, hydrogen and other metal lines are responsible for the cooling process and the gas cools to $10^{4}$~K theoretically. 

This leaves the question that whether there is an efficient cooling path from 10$^4$~K to 10$^2$~K.
An important cooling process in this cooler region is atomic structure line cooling \citep{spi78}, where permitted lines of Hydrogen cannot be excited anymore, with $<10^4$~K. This process can be observed through the FIR fine structure lines, such as [C~{\sc{ii}}] ($\lambda=158~\mathrm{\mu m}$) and [O~{\sc{i}}] ($\lambda=63~\mathrm{\mu m}$) lines.
\cite{wol95} calculated the thermal equilibrium gas temperature for the neutral interstellar medium and found that the FIR fine structure cooling is significant, if the hydrogen nuclei density reaches $n \geq 0.5$~cm$^{-3}$, then results in a rapid drop in gas temperature from $T\sim~8000$~K (typical temperature of one dominate phase, warm neutral medium: WNM) down to around 100 K (cold neutral medium: CNM) as the energy space of the [C~{\sc{ii}}] 
transition, $\Delta$E/k = 92 K, \citep{dra11}.

\subsubsection{Extended bright [O~{\sc{i}}] (63~$\um$) emission: cooling path to 100 K}

H1821 was recently observed by \textit{Herschel}/PACS \citep{jua16},
which clearly detected the [O~{\sc{i}}] 
emission line and non-detection of [C~{\sc{ii}}] line, giving the upper-bound of [C~{\sc{ii}}] 
emission line flux.
The luminosity obtained in the central spaxel, 3~$\times$~3 array, and 5~$\times$~5 array, is
$L_{\mathrm{[O~{\sc{I}}]}}$/($10^{43}$~erg~s$^{-1}$)=$43.1\pm0.5$, 76.5$\pm1.1$, 90.0$\pm2.0$ and $L_{\mathrm{[CII]}}$/($10^{43}$~erg~s$^{-1}$)~$\leq$~4.7, 10.1, 8.8 (3 $\sigma$ limit), respectively. 
The beam size for \textit{Herschel}/PACS at the band shorter than $\lambda=130$~$\mu$m is dominated by the spaxel size (field of view: 9''.4~$\sim$~42 kpc), and the resolution at 158 $\um$ is 11''.2~$\sim$~50 kpc.
This indicates that H1821 has an extended [O~{\sc{i}}] emission region larger than~42 kpc, which is consistent with the cooling radius of 90 kpc.
The [O~{\sc{iii}}] (88~$\um$) emission, which often originates from the ionized gas region produced by AGN, does not show extend feature with $L_{\mathrm{[O~{\sc{III}}]}}$/($10^{42}$~erg~s$^{-1}$)=$52.5\pm2.0$, $60.5\pm6.8$, $64.8\pm7.6$, respectively.
Thus, the extended [O~{\sc{i}}]
emission in H1821 is likely to be from the cooling flow gas.

We obtain the cooling rate through those [O~{\sc{i}}] 
emission lines as
\begin{equation}
    \begin{split}
        &\mdotcool = \frac{2}{5}\ \frac{\mu m_p  L_{\mathrm{[O~{\sc{I}}]}}}{kT} =\\
         &2.3\times 10^5\ \left(\frac{T}{10^4 \ \mathrm{K}} \right)^{-1} \left(\frac{L_{\mathrm{[O~{\sc{I}}]}}}{3.3 \times 10^{43}\ \mathrm{erg~s^{-1}}} \right)\ \msun ~\mathrm{yr^{-1}},
    \end{split}\label{equ:cool_FIR}
\end{equation}
 where we assume [O~{\sc{i}}] 
luminosity of the 3~$\times$~3 array after excluding the central spaxel (hereinafter called ``$3\times3-1$ spaxels'') originates from the cooling gas of the ICM. 
It indicates that the strong
[O~{\sc{i}}] 
emission line detected in H1821
can cool the adequate amount of gas from 10$^4$~K to 10$^2$~K. 
Therefore, H1821 already has
a reasonable cooling path from $10^7$~K to $\sim10^2$~K, which provides a large amount
of cold gas to H1821, and it would also lead to the current prompt star-formation.

\subsubsection{Origin of [O~{\sc{i}}] (63~$\um$) emission: cooling flow and/or PDRs}
The observation of [C~{\sc{ii}}] 
and [O~{\sc{i}}] 
emission lines gives
an important suggestion on the origin of these emission lines.
The non-detection of [C~{\sc{ii}}] 
gives high luminosity ratio of
$L_\mathrm{[O~{\sc{I}}]}/L_\mathrm{[C~{\sc{II}}]} >9, 7, 3 $ for the central spaxel, 3~$\times$~3 array and ``$3\times3-1$ spaxels'', respectively.
Both [C~{\sc{ii}}] and [O~{\sc{i}}] emissions are often considered to originate from the HII regions, photo-dissociation regions \citep[PDRs;][]{hol99}
outside the ionized regions of stars and X-ray dissociation regions \citep[XDRs;][]{mei07}, which are often related to the AGN activity.
The flux ratio of the two lines $L_\mathrm{[O~{\sc{I}}]}/L_\mathrm{[C~{\sc{II}}]}$ increases for some high density and/or high radiation fields in the PDRs or XDRs as analyzed in \cite{mei07}, due to the critical density and excitation temperature of the [O~{\sc{i}}]: $5 \times 10^5$~cm$^{-3}$, T$_\mathrm{ext} \sim$ 228 K and [C~{\sc{ii}}]: $3 \times 10^3$~cm$^{-3}$, T$_\mathrm{ext} \sim$ 91 K \citep{dra11}. 

Observationally, \cite{her18} found that systems with AGN show high ratio
$L_\mathrm{[O~{\sc{I}}]}/L_\mathrm{[C~{\sc{II}}]} \sim 1$ to $9$, while star-forming galaxies show one order of magnitude smaller value of
$L_\mathrm{[O~{\sc{I}}]}/L_\mathrm{[C~{\sc{II}}]} \sim 0.1$
and even ULIRGs show $L_\mathrm{[O~{\sc{I}}]}/L_\mathrm{[C~{\sc{II}}]} \lesssim 2$ at the maximum \citep[e.g.,][]{stu10}. Thus, achieving $L_\mathrm{[O~{\sc{I}}]}/L_\mathrm{[C~{\sc{II}}]} \gtrsim 3$ is in general difficult even with intensive star-formation in ULIRGs. 
In contrast, when WNM or CNM is the dominant phase in a extended region, and the gas cools from $10^4$~K to around 100~K, the ratio of radiation function in such region $\Lambda_\mathrm{[O~{\sc{I}}]}/\Lambda_\mathrm{[C~{\sc{II}}]} \sim$~a few \citep{dra11} can naturally realize $L_\mathrm{[O~{\sc{I}}]}/L_\mathrm{[C~{\sc{II}}]} \sim3$ as obtained in the extended area of H1821. 
Therefore, the value of $L_\mathrm{[O~{\sc{I}}]}/L_\mathrm{[C~{\sc{II}}]} >10$ in the central spaxel can be related to AGN activity and the value of
$L_\mathrm{[O~{\sc{I}}]}/L_\mathrm{[C~{\sc{II}}]} \gtrsim 3$ in the extended region would originate from the cooling gas of the ICM,
or extremely dense PDRs (higher than any other observed system).
This extended emission suggests that there is a star-forming region beyond the host galaxy scale, thanks to the cooling flow such as that observed in the Phoenix cluster \citep{mcd15}. 
If we assume this emission line originates from the cooling gas funneling through the cooling flow at least 1 \% of $L_\mathrm{[O~{\sc{I}}]}$ (see equation~\ref{equ:cool_FIR}), H1821 has reasonable cooling path from $10^7$~K to $10^2$~K, which provides a large amount
of cold gas to H1821, and it would also lead to the current prompt star-formation. 

\subsection{Molecular Gas: Immediate Proceeding into Star-Formation}\label{sec:Molecular}

Our results indicate that H1821 now experiences an
extreme starburst reaching $\sim 10^3~\msun$~yr$^{-1}$,
requiring a large amount of cold gas, especially
the molecular gas that proceeds the star-formation
immediately \citep[e.g.,][]{ler08,schruba11}.
\cite{ara11} detected the large CO(1-0) molecular gas blob with 4$\sigma$, which has a size of $\sim$~9~kpc, and it is located at $\sim$~9~kpc away from the central galaxy H1821.
Considering the positional accuracy 1"~$\sim$~4 kpc, the shift of $\sim9$~kpc seems a real one.
The corresponding molecular gas mass
is $8 \times 10^9\ \msun$ by using the typical CO-luminosity-to-gas-mass conversion factor for ULIRGs \citep{dow98,ara11}. 

 The counterpart of this CO gas blob is still unknown, since there is no counterpart in IR bands, neither in 2MASS and $\wise$ bands, and the optical emission is overwhelmed by the quasar emission from H1821. 
The combinations of the above results suggest that the CO gas blob is likely a stand-alone gas blob not associated with a galaxy \citep[e.g.,][]{lin09}, or there is a galaxy at the CO gas blob location, but the galaxy is completely optically and even near-IR ``dark'' because of the extreme obscurations in the host galaxy scale \citep[e.g.,][]{wan19,kub19,tob20,fud21}.
\cite{flo04} suggested that there might be a nebulous optical structure at the CO gas blog position, by applying the $\hst$ optical image decomposition from the bright quasar component at H1821. If this is true, H1821 might have a nearby galaxy at the separation of $\sim9$~kpc. It is also consistent with \cite{kal16} morphological analysis that showed the ICM of H1821 is categorized as a non-relaxed cluster. 
This arises the possibility that H1821 may have experienced a galaxy major merger with this nearby galaxy, which might have enhanced the angular momentum removal of the gas, resulting more gas supply to H1821. 

We also calculate the upper limit of the
molecular gas mass of H1821 based on the non-detection and the obtained CO(1-0) sensitivity by \cite{ara11}. Considering the 3$\sigma$ upper limit of the velocity and area integrated CO(1-0) luminosity ($S_\mathrm{CO} dv<2.1$~Jy~km~s$^{-1}$, $L^{'}_\mathrm{CO} < 4.4 \times 10^{9}$~K~km~s$^{-1}$~pc$^{2}$) from \cite{ara11}, the molecular gas mass is obtained as $M_\mathrm{H2}<3.5\times 10^{9}\ \msun$ by assuming the CO-luminosity-to-gas-mass conversion factor 0.8~$M_\odot$~(K~km~s$^{-1}$~pc$^2$)$^{-1}$ for ULIRGs \citep{dow98,ara11}. This upper limit only gives a weak constraint,
since it is still higher than the typical molecular gas mass for elliptical galaxies \citep{cro11} and also comparable to ULIRGs \citep{sol97,dow98}. Therefore, a deeper CO(1-0) observation using NOEMA (NOrthern Extended Millimeter Array \footnote{\url{https://www.iram-institute.org/EN/noema-project.php}} \footnote{\url{https://www.iram.fr/GENERAL/NOEMA-Phase-A.pdf}}) is necessary to conclude the presence of the cold gas and the cooling flow scenario at H1821. 
The CO lines with higher-J transitions or dense gas tracer (HCN, HCO+) using NOEMA can also resolve the question that the reason of CO(1-0) non-detection in H1821, such as the weakness of CO(1-0) due to the strong ionization in H1821 by starburst activity and/or AGN \citep[e.g.,][]{van10,car13, gre14,and18,izu20} or the dense PDR.
Furthermore, the high spatial resolution near-IR band imaging with the 8~m class telescopes might be able to detect the host galaxy counterpart without worrying about the contamination from the central H1821 emission. 

Summarizing the properties above,  H1821 is in a special environment where the galaxy cluster provides a large amount of cold gas which can cool down below $<10^2$~K, and such cooled gas can efficiently fall into the host galaxy thanks to the galaxy major merger with plenty of molecular gas, if it is true that there is nearby galaxy at the separation of just $\sim9$~kpc away.

\subsection{Inverse Compton Cooling is not Effective Cooling Path for H1821}\label{sec:inv}
 
 In addition to the energy balance at the galaxy cluster scale of $\sim100$~kpc, there is one cooling mechanism that might produce an additional cooling path in the host galaxy scale with $\leq10$~kpc.
 That is, inverse Compton cooling \citep{Ryb79}.
 Some studies considered that the inverse Compton cooling is responsible for increasing the flow of material from a hot atmosphere onto the black hole \citep[e.g.,][]{fab90,rus10,wal14}.
H1821 can be an ideal case
 since the accretion rate on the SMBH is high and the disk is in a radiatively efficient phase. 
However, we find inverse Compton cooling is not effective due to the X-ray emission, which originates from the hot electron corona \citep{haa91}. 
 Radiation efficiency by the inverse Compton scattering is written as 
\begin{equation}\label{equ:lambda_comp}
    \begin{split}
        \Lambda_{\mathrm{comp}} &= \frac{A}{r^2} \int d\nu L_\mathrm{AGN,\nu}(4 kT-h \nu) \\
    \end{split}
\end{equation}
 \citep{Ryb79}, where the $A=n_\mathrm{e} \sigma_\mathrm{T}/(4 \pi m_\mathrm{e} c^2)$ ($\sigma_\mathrm{T}$: cross section of Tomson scattering). Thus, the temperature ($kT$) of the cluster gas cools down to the typical temperature of AGN photon ($h\nu$). 
 In contrast, H1821 shows bright X-ray emission, and it contributes to the second term of Equation~\ref{equ:lambda_comp}. 
We calculate the equilibrium temperature ($T_\mathrm{e} =  \int d\nu h \nu L_\mathrm{AGN,\nu}/4 k L_\mathrm{AGN,bol}$, obtained from $\Lambda_{\mathrm{comp}}=0$) based on best fit model of H1821 ($L_\mathrm{AGN,\nu}$) and find that $T_\mathrm{e}=1.5$~keV. Since we assume X-ray power-law, which extend up to 300 keV and $L_\mathrm{AGN,bol}$ is two times smaller than \cite{rus10} value of $L_\mathrm{AGN,bol}=2\times 10^{47}$~erg~s$^{-1}$, this temperature is 4 times larger than \cite{rus10} value of 0.4 keV. It is comparable to the temperature of the cluster gas within host galaxy scale ($T\sim 2.4\ \mathrm{keV}$ at $\sim$~30 kpc, \citealt{rus10}). 
 Therefore, inverse Compton cooling is currently not effective at the host galaxy scale as well as the Bondi scale in H1821. 

\subsection{Possible Growth Pathways for the SMBH in H1821}
H1821 has a radiation-dominated standard disk, and it would experience the state change once the gas accretion rate drops equivalent of $\lambdaedd<10^{-2}$ \citep{nar08} at which the inner disk would be replaced by a radiatively inefficient ADAF \citep[advection-dominated accretion flow;][]{nar94,nar95} and AGN jets heat ICM. 
Assuming the current rapid accretion continue, such an accretion state change could be realized only when the BH mass reaches one order of magnitude higher, i.e., $\mbh \sim 6 \times 10^{10} \ \msun$ and thus the corresponding $\lambdaedd$ reaches $\lambdaedd \sim 10^{-2}$ in around $4\times 10^9$~yr. 

In reality, a large influx of the cooling flow to the SMBH could produce a nuclear starburst as suggested by several studies \citep[e.g.,][]{tho05,ina16}, which would start reducing $\mdotbh$ at some point. Furthermore, radiation feedback by the quasar itself can reduce the accretion rate by driving out gas out of the central region \citep{cre03,di05}. Actually, most SDSS quasars show AGN jets before reaching the maximum mass end, notably starting from $\mbh \sim 2$--$3\times 10^9 \msun$ \citep{ich17b}. 
In these cases, the rapid growth phase is expected to last only $\sim 10^{8}$~yr, which is a few dynamical timescales that takes into account galaxy mergers as described in Section \ref{sec:Molecular} \citep[and see][]{hop10,ina16} or typical quasar life time from ionized gas observation \citep{khr21} and galaxy merger simulations \citep{di05}. 
Thus, the mass of SMBH can increase efficiently until $\mbh \sim 6 \times 10^{9}\ \msun$ only in $\sim 10^{8}$~yr, then the accretion rate drastically decreases and producing strong jets with large amount of energy, $W_{\mathrm{heat, jet}}$, preventing the ICM cooling and star-formation as shown in most BCGs in the host galaxy clusters \citep{pet03,fab12}. 

\section{conclusion}\label{conclusion}

H1821+643 (hereafter, H1821) is the most luminous non-blazer AGN in the 14--150~keV band at $z<0.4$ detected in the \textit{Swift}/BAT hard X-ray all-sky survey, and it is also the most IR luminous source among the \textit{Swift}/BAT survey. Considering that H1821
is located in the center 
of a massive galaxy cluster hosting a cool core, 
it is the 
best laboratory to study how the central SMBH and the host galaxy assemble the mass in this extreme environment.

We have collected the broad-band SED covering from the X-ray to FIR of H1821 and
conducted the SED fitting by using the \verb|CIGALE| code, which enables us to estimate the gas consumption rate as the sum of the SFR and BH accretion rate $\mdotbh$, 
and we have discussed the origin of the cold gas.
Our results are summarized as follows.

\begin{enumerate}
\item H1821 consumes a large amount of gas with 
$\log ( \mathrm{SFR}/\msun~\mathrm{yr}^{-1}) = 3.01 \pm 0.04$ and \\
$\log (\mdotbh/\msun~\mathrm{yr}^{-1}) = 1.20 \pm 0.05$. 

\item This high gas consumption rate is higher than the gas cooling rate of the ICM, $\dot{M}_\mathrm{cool}= (3 -5)\times 10^2 \ \msun$~yr$^{-1}$ \citep{rus10,mcd18}, within the cooling radius of 90~kpc. This high gas consumption efficiency ($\dot{M}_\mathrm{con}/\dot{M}_\mathrm{cool} \gtrsim 1$) is one to two order magnitude higher than the typical value of other cool-core clusters. 

\item The energy balance between the cooling (by ICM X-ray radiation) and heating (by AGN jet) in the H1821 system indicates that (ICM cooling) / (AGN jet heating) $>$~1.
In addition, BH in H1821 has the radiation-efficient accretion disk with $\eddington \sim 0.2$. Even thought there is a X-ray cavity in H1821, those results may create an environment where the ICM gas can be cooled on average without interruption by heating of the jets.

\item A large fraction of the current rapid gas consumption could be provided through the ICM cooling flow, if there is a path for the gas to cool down below 100~K. 
We show that H1821 has an efficient cooling path achieving from $10^7$~K to $10^2$~K. H1821 has a bright FIR fine structure emission [O~{\sc{i}}] 63 $\um$ ($L_{\mathrm{[O~{\sc{I}}]}}=3.3 \times 10^{43}$~erg~s$^{-1}$) extends over 40 kpc, which is a main coolant in the low temperature range ($10^4$~K to $10^2$~K). 
This emission corresponds to a cooling rate of $\mdotcool =2.3\times 10^5\ \mathrm{\msun ~yr^{-1}}$. 
Thus, the X-ray gas ($\sim$~0.01 cm$^{-3}$)  in H1821 is expected to cool with $P=nkT$, resulting the decreasing of the volume (filamentation) and thus the increasing of the gas density. Once the gas density reaches $n>$~0.5~cm$^{-3}$, the cooled gas starts to emit the [O~{\sc{i}}] line, producing an effective cooling path to 100~K and finally the extended star-forming regions. 

\item H1821 is also known to have a large amount of molecular gas ($8 \times 10^9\ \msun$), which proceeds the star-formation immediately. 
In contrast, the CO (1-0) gas offset from the optical center indicates that H1821 has been experiencing the galaxy major merger, enhancing the cooled ICM gas accretion to the host galaxy and the central SMBH. 

\item If there is no negative feedback to regulate the accretion rate to the BH, this rapid growing phase for both the central BH and the host galaxy would continue until the BH reach $6 \times 10^{10}\ \msun$ with $\eddington \leq 10^{-2}$ in $4 \times 10^9$~yr, where the disk will experience the phase transition into the ADAF state.   

\item We examine the possible growth path ways of the central SMBH with reduced BH accretion rate. 
If the gas is depleted near 100~pc by star-formation suggested by \cite{ina16} or if the quasar feedback removes the gas from the center, the BH would grow to $\sim 6 \times 10^{9}\ \msun$ in $\sim 10^8$~yr
until the accretion disk changes into the ADAF state. Then, large-scale AGN jets would suppress the ICM cooling and star-formation activity in the host galaxy.

\end{enumerate}

\acknowledgments

We thank the anonymous referee for helpful suggestions that strengthened the paper.
This work is supported by Program for Establishing a Consortium
for the Development of Human Resources in Science
and Technology, Japan Science and Technology Agency (JST) and is partially supported by Japan Society for the Promotion of Science (JSPS) KAKENHI (20H01939; K.~Ichikawa). 
We acknowledge the support of Fondecyt Iniciacion grant 11190831 and the ANID BASAL project FB210003 (CR).

This publication makes use of data products from the Two Micron All Sky Survey, which is a joint project of the University of Massachusetts and the Infrared Processing and Analysis Center/California Institute of Technology, funded by the National Aeronautics and Space Administration and the National Science Foundation \citep{2mass}.
This publication makes use of data products from the Wide-field Infrared Survey Explorer, which is a joint project of the University of California, Los Angeles, and the Jet Propulsion Laboratory/California Institute of Technology, funded by the National Aeronautics and Space Administration~\citep{wise}.
This work is in part on observations made with the $\spitzer$ Space Telescope, which was operated by the Jet Propulsion Laboratory, California Institute of Technology under a contract with NASA.
This dataset or service is made available by the Infrared Science Archive (IRSA) at IPAC, which is operated by the California Institute of Technology under contract with the National Aeronautics and Space Administration~\citep{akari_irc,akari_fis}.



\appendix

\section{Multi-wavelength Properties of H1821+643}\label{app:h1821}
The multi-wavelength characteristics and past observations of H1821 are summarized here.
H1821 was identified as the center galaxy of a galaxy cluster at $z=0.297$, firstly in the optical \citep{sch92}, and then followed by 
the X-ray observations, first obtained by \textit{ROSAT}/HRI \citep{hal97,sax97}
and consectively observed by several X-ray satellites (\textit{ROSAT} PSPC: \citealt{sax97,sax99}, Chandra: \citealt{fan02,rus10}, XMM-{\sl Newton}: \cite{rui07} and \textit{Swift}/BAT survey: \citealt{ric17}).

Optical spectrum was obtained multiple times \citep[e.g.,][]{kol93},
and \cite{kos17} measured the H$\beta$ based BH mass of $\mbh = 3.9 \times 10^{9} \ \msun$
as a part of the BASS Survey. 
H1821 is known to contain a small radio jet \citep{blu96,blu01} and its heating effect on the interstellar medium is discussed in chapter \ref{sec:dis}. \cite{rus10} suggested that the small radio emission and a corresponding X-ray cavity is related to the transition phase from low/hard (radio mode AGN) to high/soft state (quasar mode) as an analogy of X-ray binaries.
Thus, this AGN jet may not be related to the typical radio-mode feedback, but can be explained by the transition to the quasar-mode \citep{fen04}. 

\section{SED fitting}\label{app:sedfit}
\subsection{Analysis of ISO Spectra}\label{app:spi}
SED fitting cannot be performed simultaneously with photometric data in \verb|CIGALE|, so we averaged the $\iso$ spectrum as we mentioned in section \ref{sec:spi}

\begin{figure*}
\begin{center}
\includegraphics[width=0.46\textwidth]{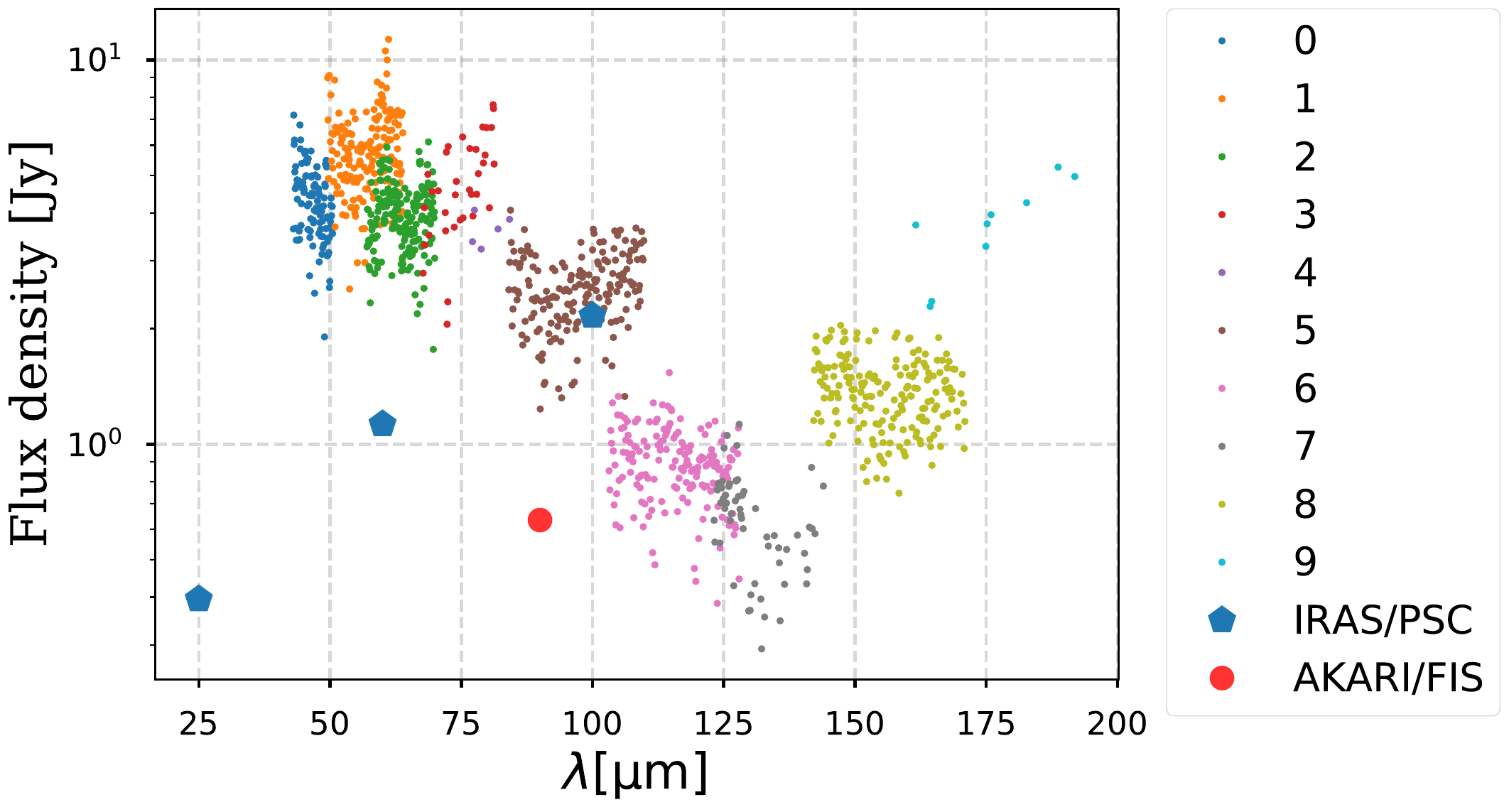}~
\includegraphics[width=0.50\textwidth]{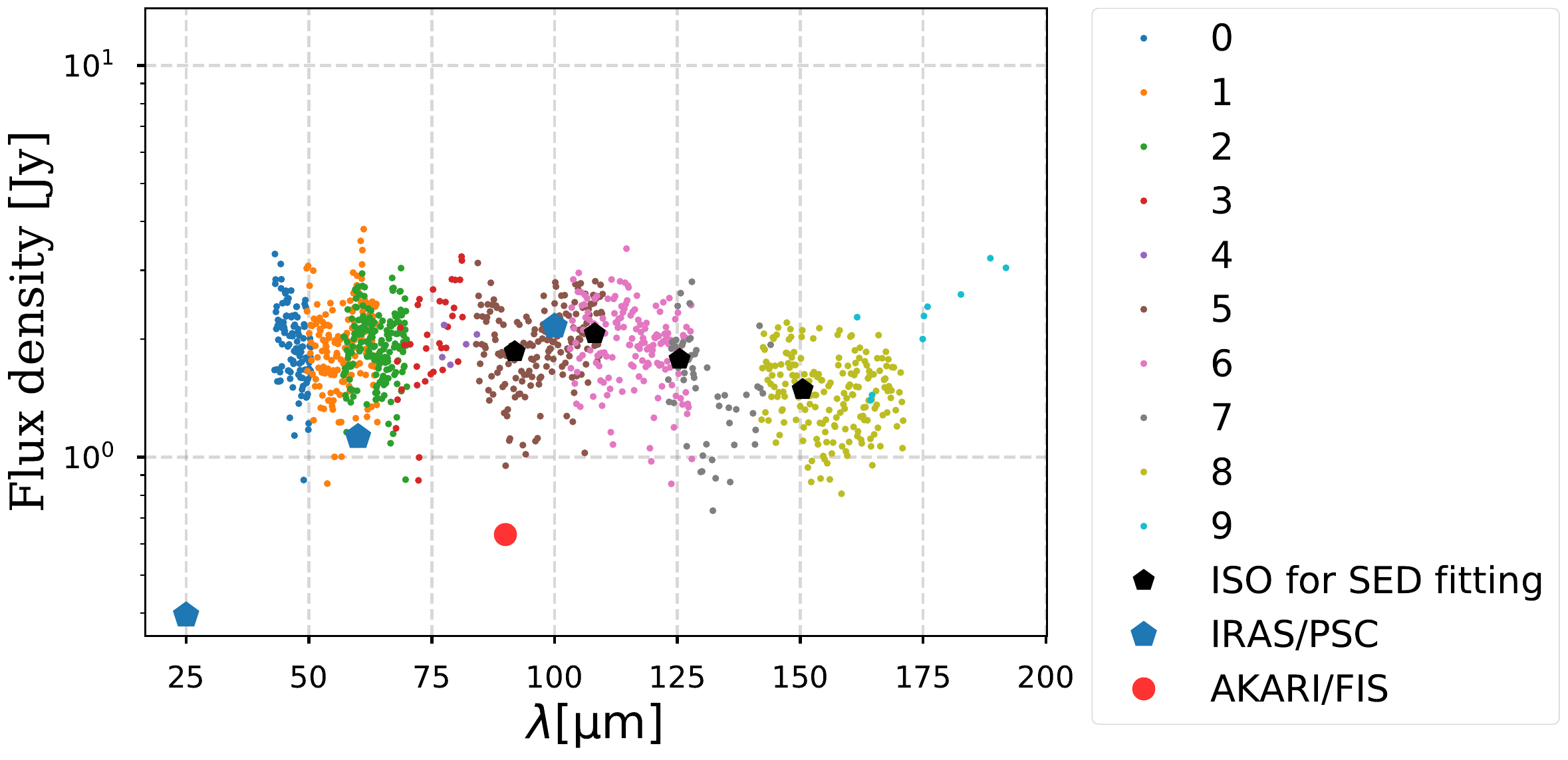}~
\caption{
(Left) ISO spectra before calibration with S/N $>$~3. 
(Right) ISO spectra after calibration using detector 5, whose flux is consistent with the  $\iras$/PSC 100 $\um$ band flux. The average spectra of the four bins are over plotted with black circles.
}\label{fig:iso}
\end{center}
\end{figure*}

The $\iso$ data was first calibrated for each detector based on the $\iras$/PSC 100 $\um$ values of 1.94~Jy, after the subtraction of the estimated PN flux (Section \ref{sec:iras}), and the corresponding detectors (number 5), since the calibration was different for each detector as shown in Figure \ref{fig:iso} (left). 

The corrected spectrum is shown in the right panel of Figure \ref{fig:iso}. 
After the calibration, we averaged the $\iso$ spectrum in the wavelength range of 100 to 180 $\um$ by dividing into four parts for tracing the emission of the host galaxy dust.

\subsection{Effect of Error Size and the Select of FIR Data on SED Fitting }\label{app:err}

\verb|CIGALE| implements the 10\% systematic error for all photometric points as default manner to avoid the over-fitting in the NIR-MIR bands,
which generally has considerably small error-bars ($<$1\%) compared to FIR ones ($\sim$10\%). 
\verb|CIGALE| code uses error values of $\sqrt{\mathrm{photo_{err}^2+0.1\times flux}}$, where the $\mathrm{photo_{err}}$ is photometric error value of each data point. 
In the case of H1821, another concern is a possible FIR contamination from nearby sources, which may cause an overestimation of the SFR. 
Thus, we investigate the effect of error size on SED fitting by changing
$\mathrm{photo_{err}}$ between the X-ray and FIR bands from the observational values to 20\% of their fluxes.
Besides, we check the result of SED fitting by using $\akari$/FIS flux instead of $\iras$/PSC flux, it is summarized in Table~\ref{tab:ape_err}, and Figure~\ref{fig:sed_err} is the best SED fitting results for each situation. 
When we change the error sizes of photometric data, the physical values such as SFR, stellar mass, and AGN luminosity do not change much. It indicates that the error size of 10\% systematic error is enough for the SED fitting, and effect of the additional photometric error and the FIR contamination is negligible.
If we use $\akari$/FIS flux instead of $\iras$/PSC flux, the FIR range of the best fitting result trace $\akari$/FIS flux, resulting in slightly smaller SFR by a factor of 2 ($\sim 600 \ \msun$~yr$^{-1}$).

 Therefore, although SFR becomes a change by factor of a few for the case with $\akari$/FIS flux, we concluded that the set of error size and FIR normalization does not affect our study. 
 
\begin{deluxetable*}{cccccccc}
\tabletypesize{\footnotesize}
\tablecolumns{8}
\tablewidth{0pt}
\tablecaption{Result of SED fitting \label{tab:ape_err}}
  \tablehead{
  \colhead{Photometric error size}&
      \colhead{Reduced chi square} &\colhead{$L_{\mathrm{AGN,IR}}$} &\colhead{$L_{\mathrm{AGN,bol}}$} 
      &\colhead{$\mdotbh$}
      & \colhead{$L_{\mathrm{host,IR}}$} &\colhead{$\mstar$} & 
\colhead{SFR} 
      \\
      \colhead{} &
       \colhead{} & \colhead{($10^{46}$~erg~s$^{-1}$)} & \colhead{($10^{46}$~erg~s$^{-1}$)}
       & \colhead{($\msun$~yr$^{-1}$)} &\colhead{($10^{46}$~erg~s$^{-1}$)} & \colhead{($10^{11}$~$\msun$)} & 
       \colhead{($10^{2}$~$\msun$~yr$^{-1}$)}
     }
\startdata
 \multicolumn{8}{c}{Using $\iras$ and ISO data}\\
photometric &                     1.0 &  46.51 $\pm$ 0.04 &  46.96 $\pm$ 0.05 &  1.20 $\pm$ 0.05 &  46.35 $\pm$ 0.04 &  11.57 $\pm$ 0.38 &  3.01 $\pm$ 0.04 \\
5\%     &                     1.0 &  46.51 $\pm$ 0.04 &  46.95 $\pm$ 0.05 &  1.20 $\pm$ 0.05 &  46.35 $\pm$ 0.04 &  11.56 $\pm$ 0.38 &  3.01 $\pm$ 0.04 \\
10\%    &                     0.8 &  46.53 $\pm$ 0.05 &  46.97 $\pm$ 0.05 &  1.22 $\pm$ 0.05 &  46.36 $\pm$ 0.05 &  11.65 $\pm$ 0.39 &  3.02 $\pm$ 0.05 \\
20\%    &                     0.3 &  46.53 $\pm$ 0.06 &  46.98 $\pm$ 0.06 &  1.23 $\pm$ 0.06 &  46.36 $\pm$ 0.07 &  11.64 $\pm$ 0.41 &  3.02 $\pm$ 0.07 \\
 \hline         
 \multicolumn{8}{c}{Using $\akari$ data}\\
photometric &                     1.6 &  46.49 $\pm$ 0.02 &  46.95 $\pm$ 0.03 &  1.19 $\pm$ 0.03 &  46.12 $\pm$ 0.02 &  11.53 $\pm$ 0.41 &  2.78 $\pm$ 0.02 \\
5\%     &                     1.6 &  46.49 $\pm$ 0.02 &  46.95 $\pm$ 0.03 &  1.19 $\pm$ 0.03 &  46.12 $\pm$ 0.02 &  11.51 $\pm$ 0.40 &  2.77 $\pm$ 0.02 \\
10\%    &                     1.0 &  46.47 $\pm$ 0.02 &  46.93 $\pm$ 0.03 &  1.18 $\pm$ 0.03 &  46.10 $\pm$ 0.02 &  11.49 $\pm$ 0.41 &  2.76 $\pm$ 0.02 \\
20\%    &                     0.4 &  46.45 $\pm$ 0.03 &  46.92 $\pm$ 0.05 &  1.17 $\pm$ 0.05 &  46.08 $\pm$ 0.03 &  11.52 $\pm$ 0.45 &  2.74 $\pm$ 0.03 \\
\enddata
\tablecomments{Physical quantities obtained by SED fitting for each FIR data and each error size. Error sizes are photometric error from observed error, 5\% flux, 10\% to 20\% plus 10\% systematic error, which is set as default value in CIGALE-2022.0
}
\end{deluxetable*}

\begin{figure*}
\begin{center}
\includegraphics[width=0.98\textwidth]{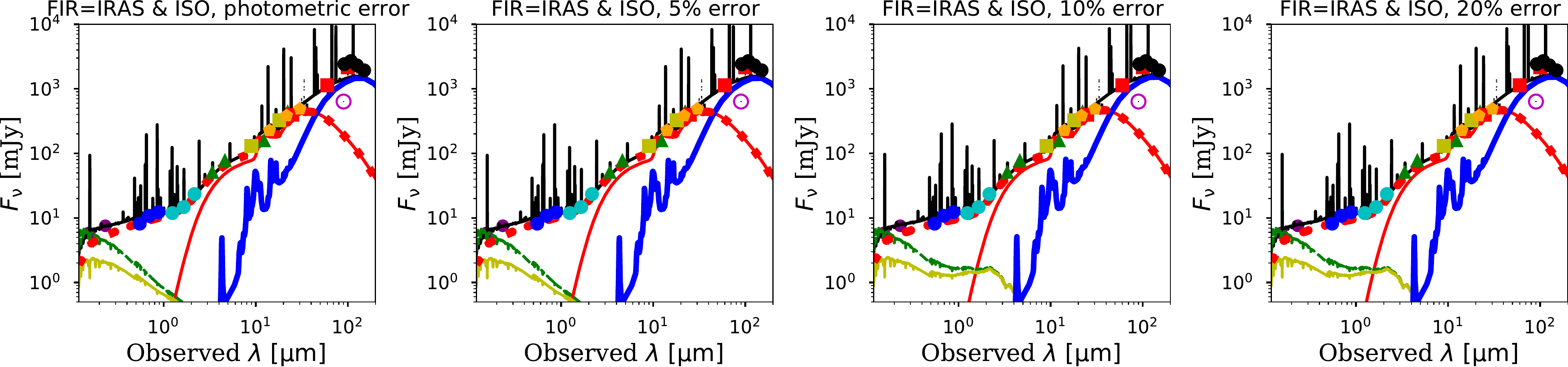}~
\caption{Best SED fitting results of H1821 for the case of changing the error size with using $\iras$ and ISO fluxes for FIR data. The filled points are the photometric data and the symbols are the same as in Figure~\ref{fig:1821sed}. 
}\label{fig:sed_err}
\end{center}
\end{figure*}
\begin{figure*}
\begin{center}
\includegraphics[width=0.98\textwidth]{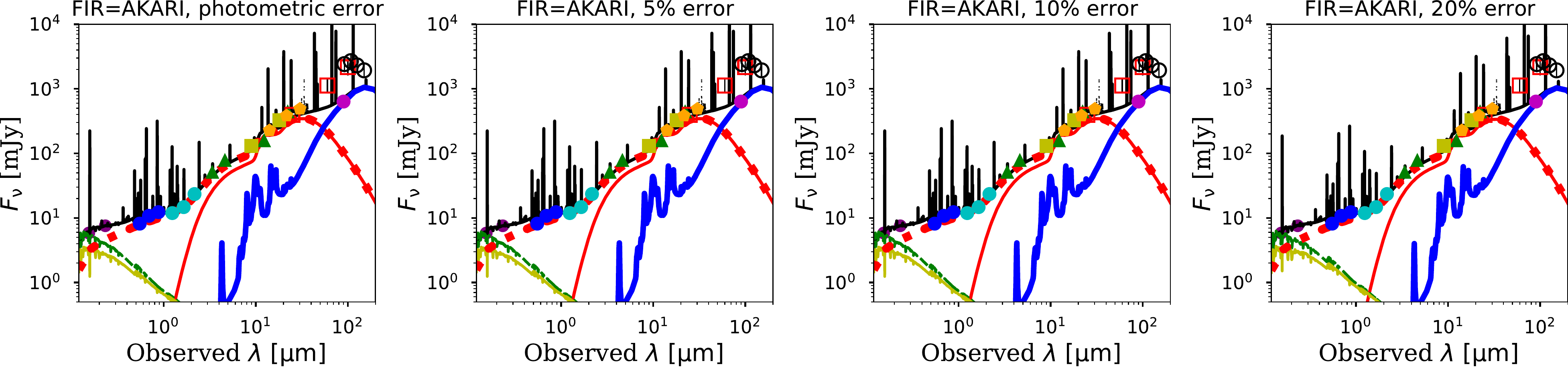}~
\caption{Same with Figure~\ref{fig:sed_err} but using $\akari$ flux for FIR data. 
}\label{fig:sed_err}
\end{center}
\end{figure*}



\pagebreak

\bibliographystyle{aasjournal}
\bibliography{bibtex}



\end{document}